\documentclass[10pt, letterpaper, notitlepage, oneside]{article}

\usepackage[
	letterpaper,
	bottom      = 1.00in,
	left        = 0.75in,
	right       = 0.75in,
	top         = 1.00in
]{geometry}
\usepackage[utf8]{inputenc}
\usepackage[nopar]{lipsum}

\usepackage{fancyhdr}
\fancypagestyle{specialfooter}{
	\fancyhf{}
	\lhead{Published in \textit{Physical Review Fluids}: \href{https://doi.org/10.1103/PhysRevFluids.3.113101}{\url{https://doi.org/10.1103/PhysRevFluids.3.113101}} \hfill \thepage}
	\cfoot{$\dag$ \textbf{Email address for correspondence:} g\_dilabb{\MVAt}encs.concordia.ca}
}

\usepackage{hyperref}
\hypersetup{
	breaklinks = true,
	citecolor  = blue,
	colorlinks = true,
	filecolor  = blue,
	linkcolor  = blue,      
	urlcolor   = blue
}

\usepackage{titlesec}
\titleformat*{\section}{\Large\bfseries}
\titleformat*{\subsection}{\normalsize\bfseries\filcenter}
\titleformat*{\subsubsection}{\normalsize\bfseries}

\makeatletter
\renewcommand\@seccntformat[1]{\csname the#1\endcsname.\quad}
\makeatother

\usepackage{caption}
\usepackage{cleveref}
\usepackage{graphicx}
\usepackage{epstopdf}
\usepackage[
	font = normalsize
]{subfig}
\usepackage{xparse}
\crefformat{subfigure}{#2\onlyletter{#1}#3}
\crefrangeformat{subfigure}{#1--#3\onlyletter{#2}#4}
\ExplSyntaxOn
\NewDocumentCommand{\onlyletter}{m}{
	\tl_set:Nx  \l_tmpa_tl { #1 }
	\tl_item:Nn \l_tmpa_tl { -1 }
}
\ExplSyntaxOff

\usepackage{amsfonts}
\usepackage{amsmath}
\usepackage{marvosym}
\usepackage{textcomp}
\usepackage{upgreek}
\newcommand{\volume}{{\ooalign{\hfil$V$\hfil\cr\kern0.08em--\hfil\cr}}}

\usepackage{arydshln} % For creating dashed lines in tables.
\usepackage{multirow}
\usepackage{setspace}

\usepackage{apacite} % APA Style
\let\cite\shortcite
\let\citeA\shortciteA

\newcommand{\noop}[1]{}

\begin{document}
	
	% First Page Footer.
	\thispagestyle{specialfooter}
	
	% Title.
	\noindent
	\textbf{\LARGE Material transport in the left ventricle with aortic valve\\\\ regurgitation}\hfill\break
	
	% Abstract.
	\begin{changemargin}{0.25in}{0.00in}
		{\large Di Labbio G$^{1,2\dag}$, V\'{e}tel J$^{2}$, Kadem L$^{1}$}\hfill\break
		$^{1}$\textit{Laboratory of Cardiovascular Fluid Dynamics, Concordia University, Montr\'{e}al, QC, Canada, H3G 1M8}\hfill\break
		$^{2}$\textit{Laboratoire de Dynamiques des Fluides, Polytechnique Montr\'{e}al, Montr\'{e}al, QC, Canada, H3T 1J4}\hfill\break
		
		% Write here (remove lipsum line below).
		This experimental \textit{in vitro} work investigates material transport properties in a model left ventricle in the case of aortic regurgitation, a valvular disease characterized by a leaking aortic valve and consequently double-jet filling within the elastic left ventricular geometry. This study suggests that material transport phenomena are strongly determined by the motion of the counterrotating vortices driven by the regurgitant aortic and mitral jets. The overall particle residence time appears to be significantly longer with moderate aortic regurgitation, attributed to the dynamics resulting from the timing between the onset of the two jets. Increasing regurgitation severity is shown to be associated with higher frequencies in the time-frequency spectra of the velocity signals at various points in the flow, suggesting nonlaminar mixing past moderate regurgitation. Additionally, a large part of the regurgitant inflow is retained for at least one cardiac cycle. Such an increase in particle residence time accompanied by the occurrence and persistence of a number of attracting Lagrangian coherent structures presents favorable conditions and locations for activated platelets to agglomerate within the left ventricle, potentially posing an additional risk factor for patients with aortic regurgitation.
		\hfill\break
		\textbf{PhySH:} \textit{Cardiac haemodynamics}; \textit{Heart diseases}; \textit{Pathology}; \textit{Physiological flows}; \textit{Vortex flows}
		
		\hfill\break
		* \textit{Data pertaining to this article will be made available from the authors upon reasonable request.}\hfill\break
		* \textit{The authors have no conflicts of interest to declare.}\hfill\break
		* \textit{Please cite as}: Di Labbio, G., V\'{e}tel, J., \& Kadem, L. (2018). Material transport in the left ventricle with aortic valve regurgitation. \textit{Physical Review Fluids}, \textit{3}(11), 113101.
		
		\hfill\break
		\raisebox{0.95pt}{\small\textcopyright} 2018 American Physical Society. This manuscript version is made available under the APS Copyright Transfer Agreement (\href{https://journals.aps.org/authors/transfer-of-copyright-agreement}{\url{https://journals.aps.org/authors/transfer-of-copyright-agreement}}), more information regarding usage terms can be found at \href{https://journals.aps.org/info/terms.html}{\url{https://journals.aps.org/info/terms.html}}. The published version of the manuscript is available at \href{https://doi.org/10.1103/PhysRevFluids.3.113101}{\url{https://doi.org/10.1103/PhysRevFluids.3.113101}}. The supplementary information associated with this manuscript can be found along with the published article at the same link.
	\end{changemargin}
	
	\section{\label{sec:Intro}Introduction}
	
		Cardiovascular diseases are still comprehensively the leading cause of death in the world, killing as many as $17.7$ million people in $2015$, equivalent to $31$\% of all global deaths that year \cite{WHO17}. Evidently, there is a vast amount of research aiming to understand the etiology and subsequent pathophysiology of cardiovascular diseases in the hopes of offering more effective means of detection, ultimately promoting early treatment measures and preventive medicine. With the heart being such an intricate periodic fluid pump, it comes as no surprise that the fluid dynamics community have found a niche in studying disease etiology and pathophysiology in terms of blood flow dynamics with the same ambitions over the past two decades. The kernel for many of these studies was the beautiful magnetic resonance flow visualizations of \citeA{Kilner00} and their insightful hypothesis that the natural swirling flows in the healthy heart chambers minimize energy dissipation. Subsequent studies have focused mainly on an Eulerian flow description in the left ventricle, the largest and most laborious chamber, as its diseases are comparatively more debilitating than those of other heart chambers. The reader is referred to the work of \citeA{PedrizzettiDomenichini15} for a thorough review of modern developments in left ventricular fluid mechanics. Although there is certainly much to gain from an Eulerian analysis of the flow in the left ventricle, blood is ultimately a suspension of particles comprised of platelets and both red and white blood cells, giving a rather powerful purpose to analyzing the flow from a Lagrangian perspective.
		
		In view of the limited information available on actual material transport in the left ventricle, by tracking virtual material elements, \citeA{Bolger07} sought to compartmentalize the left ventricular blood volume into four distinct components, namely, that which ($1$) has entered and ejected the ventricle in the same beat (direct flow), ($2$) has entered and remained in the ventricle in the current beat (retained inflow), ($3$) was already present in the ventricle and was ejected in the current beat (delayed ejection flow), and ($4$) was already present in the ventricle and not ejected in the current beat (residual volume). In this manner, the overall pumping efficiency of the left ventricle, in a Lagrangian sense, is effectively described on a per cycle basis. Their interest in material volumes revealed that particles following the direct flow path conserve more of their kinetic energy. In the case of a dilated and weakened left ventricle, a disease known as dilated cardiomyopathy, they observed a reduction in the volume fraction of direct flow in a single patient, which was further supported in a small cohort of patients in \citeA{Eriksson13}, suggesting an overall lower kinetic-energy-conserving efficiency and longer residence time of blood cells in the left ventricle. Other than direct particle advection, Lagrangian coherent structures (LCSs) have also found use in cardiovascular flows through the works of \citeA{ShaddenTaylor08}, \citeA{Shadden08}, \citeA{Vetel09} and \citeA{Xu09}. Lagrangian coherent structures are material surfaces (or curves in two dimensions) that act as organizing centers for particle advection patterns, the full variational theory and computation of which could be found in \citeA{Haller11} and \citeA{Farazmand12, Farazmand12errat} (the reader is referred to the work of \citeA{Haller15} and the references therein for a comprehensive review). Although both false positives and negatives may arise in using ridges of the finite-time Lyapunov exponent (FTLE) field to characterize LCSs \cite{Haller11}, the FTLE field remains a rather simple heuristic diagnostic for observing LCSs and has been the method of choice in the cardiovascular flow community. Ridges of the FTLE field still nearly maintain the property of being material transport barriers, having negligible flux across them \cite{Shadden05} (under the revised assumptions of \citeA{Haller11}), and are therefore useful in delineating blood volumes separated by their dynamics. With regard to flow in the left ventricle, \citeA{Toger11} first showed the potential of LCSs to distinguish blood entering the ventricle from blood already present in the ventricle. They then used LCSs to quantify the volume of the vortex ring developing during left ventricular filling in \citeA{Toger12b,Toger12a} from \textit{in vivo} magnetic resonance velocity data, showing little difference in the vortex volume between healthy subjects and patients with dilated cardiomyopathy; although the authors stress that the vortex volume fraction is considerably lower in the patients, this is simply a consequence of the larger ventricle volume characteristic of dilated cardiomyopathy. \citeA{Charonko13} demonstrated the importance of the mitral valve leaflets in left ventricular filling, where LCSs extend downstream from the leaflets to form a channel, guiding the mitral inflow jet toward the ventricle apex and delaying its expansion. From \textit{in vivo} Doppler echocardiography velocity data, benefitting from the higher temporal resolution over that of magnetic resonance imaging, \citeA{HendabadidelAlamoShadden12}, \citeA{Hendabadi12} and \citeA{Hendabadi13} used LCSs to reveal blood transport and residence time behavior throughout the left ventricle with elegant illustrations in a small cohort of both healthy subjects and patients with dilated cardiomyopathy. Their use of attracting and repelling LCSs computed over two cardiac cycles to respectively delineate injected and ejected blood volumes enabled them to redefine the aforementioned compartmentalized volumes of \citeA{Bolger07} in terms of intersections of these LCS-delineated regions. The authors additionally partitioned the left ventricular volume further based on residence time in a Lagrangian sense. Rather importantly, they also showed that Eulerian measures of particle residence time, although computationally inexpensive, underestimate what is suggested by a Lagrangian analysis.
		
		In general, the inherent lack of control and repeatability in performing \textit{in vivo} studies makes it rather difficult to investigate the incremental effects of some conditions or pathologies in isolation, often requiring tedious data collection from a considerably large number of patients and the use of statistical inferences to judge the significance of the reported correlations. It is more effective to investigate pathophysiology with \textit{in vitro} or \textit{in silico} models and then validate the findings against \textit{in vivo} data. In keeping with a Lagrangian analysis of left ventricular flow, from an \textit{in vitro} perspective, \citeA{Espa12} computed LCSs in a simulated healthy left ventricle, where they described the evolution of the diastolic vortex in effectively two orthogonal planes and later in three dimensions in \citeA{Badas13, Badas15}, showing qualitative agreement with the discussed \textit{in vivo} results. Highly resolved surfaces of attracting and repelling LCSs have recently been extracted in the three-dimensional numerical simulations of \citeA{Badas17}, showing quite remarkably the evolution of the diastolic vortex ring for a healthy left ventricle and one with part of the ventricular wall being dyskinetic (known as myocardial infarction). The attracting LCS representing the vortex ring in the case of the infarcted left ventricle remained rather localized near the mitral valve, suggesting a large quiescent zone within the ventricle characterized by long particle residence time and offering some degree of explanation for the development of mural thrombi in such patients. Lagrangian coherent structures in \textit{in vitro} experiments have also been used to investigate the effects of a left ventricular assist device in a model left ventricle with dilated cardiomyopathy \cite{Herold16, MayNewman16} and to study the intraventricular flow and performance of prosthetic valves \cite{Herold16, Wang17}.
		
		\begin{figure}[!t]
			\centering
			\includegraphics[width=0.45\linewidth]{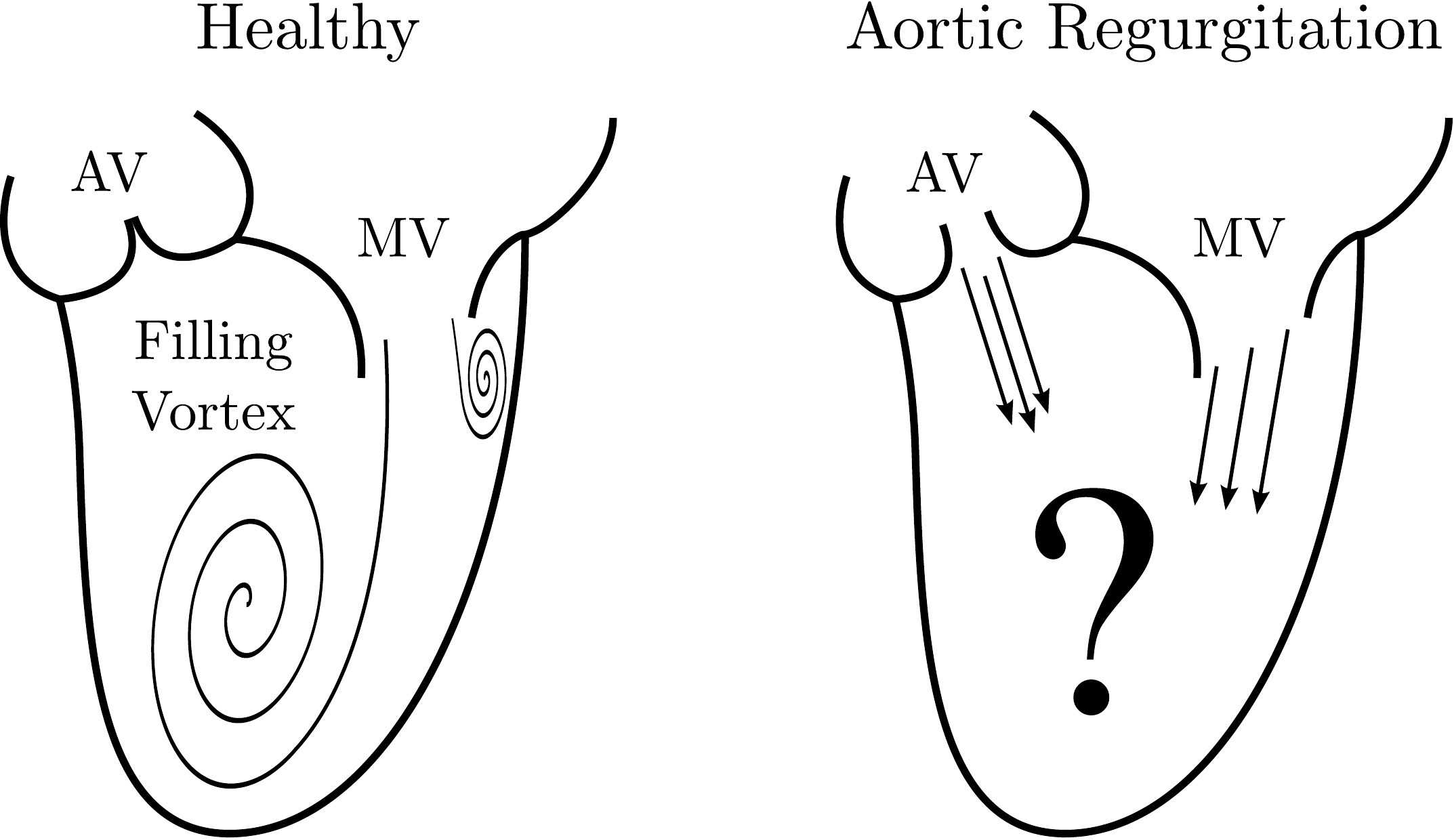}
		\caption{Schematic of the filling flow in a healthy left ventricle and one with aortic valve regurgitation. Here and in the following figures AV denotes aortic valve, MV mitral valve, and ROA regurgitant orifice area.}
		\label{fig:LVschem}
		\end{figure}
		
		As may have been inferred by the present discussion, significant research effort has been allocated to studying the intraventricular fluid dynamics particularly associated with dilated cardiomyopathy. Very few studies have investigated intraventricular flow in the case of other prevalent pathologies of the left ventricle. In particular, valvular diseases present themselves as quite the common pathology, increasing exponentially in prevalence with age \cite{Nkomo06, Iung11, Coffey16}. With the average life expectancy consistently on the rise, valvular heart diseases are believed to be the next cardiac epidemic of this century \cite{dArcy11}. As such, this work focuses on a specific valvular disease with intriguing yet elusive underlying fluid dynamics, namely, aortic regurgitation. Aortic regurgitation, also known as aortic insufficiency, is a condition where left ventricular filling partly occurs from leakage through the aortic valve and is rather common, having a reported prevalence of $5.2$\% in the adult population of $40$ years or greater in the USA \cite{Singh99}. This leakage prompts an interaction between two pulsatile jets, namely, that from regurgitation and that from natural mitral inflow, in a confined elastic geometry (see Fig.\ \ref{fig:LVschem} for a schematic). In the case of chronic (gradually occurring) aortic regurgitation, the left ventricle is known to dilate in an attempt to accommodate the regurgitant volume while maintaining a healthy forward stroke volume to meet the body's oxygen and nutrient demands \cite{Bekeredjian05, Stout09}. The continual volume overload imposed on the left ventricle by the regurgitation and its consequent dilation can impair proper functioning of the left ventricle, causing dilated-cardiomyopathy-like features with the added fluid dynamic complication of the regurgitant aortic jet. In a previous study, we have shown that the corresponding intraventricular flow results in increased viscous energy loss and gradual vortex reversal with regurgitation severity from \textit{in vitro} simulations \cite{Raymondet16, DiLabbio16, BenAssa17, DiLabbioKadem18}. The increased energy loss has also been previously observed \textit{in vivo} by \citeA{Stugaard15} in a small cohort of patients with chronic aortic regurgitation, however the methodology used to capture their velocity fields is known to have significant limitations \cite{PedrizzettiSengupta15} and thus there remains a need for a comprehensive \textit{in vivo} investigation of the resulting intraventricular flow. The \textit{in vitro} work of \citeA{Okafor17} also demonstrated this increased viscous energy loss, however they did not reproduce the physiological adaptations seen in chronic aortic regurgitation, limiting their study to acute (suddenly occurring) cases at best. Additionally, by nature, it is the relaxation of the heart muscle that induces filling from both the aortic and mitral valves and so, contrary to their work, little to no regurgitation should be expected in the late filling period during which ejection of the left atrium is responsible for left ventricular filling. This lack of independent ventricular and atrial activation in their system is a critical feature in experimentally investigating a disease like aortic regurgitation. Nevertheless, their results do suggest that a kinematic barrier to mitral inflow may develop as a consequence of aortic regurgitation, in their case due to the regurgitant jet, although prolonged, impacting the opposite ventricular wall and in the case of \citeA{DiLabbioKadem18} (and the present study) due to the development of a counterrotating vortex driven by the regurgitant jet. How the resulting intraventricular flows organize material for ejection and the corresponding blood stasis tendencies remain to be investigated. Here the material transport characteristics of the flow in the left ventricle subject to aortic regurgitation will be investigated \textit{in vitro}, taking into account the physiological adaptations occurring when the regurgitation is chronic as well as the independence of atrial and ventricular activation.
		
		\begin{figure}[!t]
			\centering
			\includegraphics[width=0.50\linewidth]{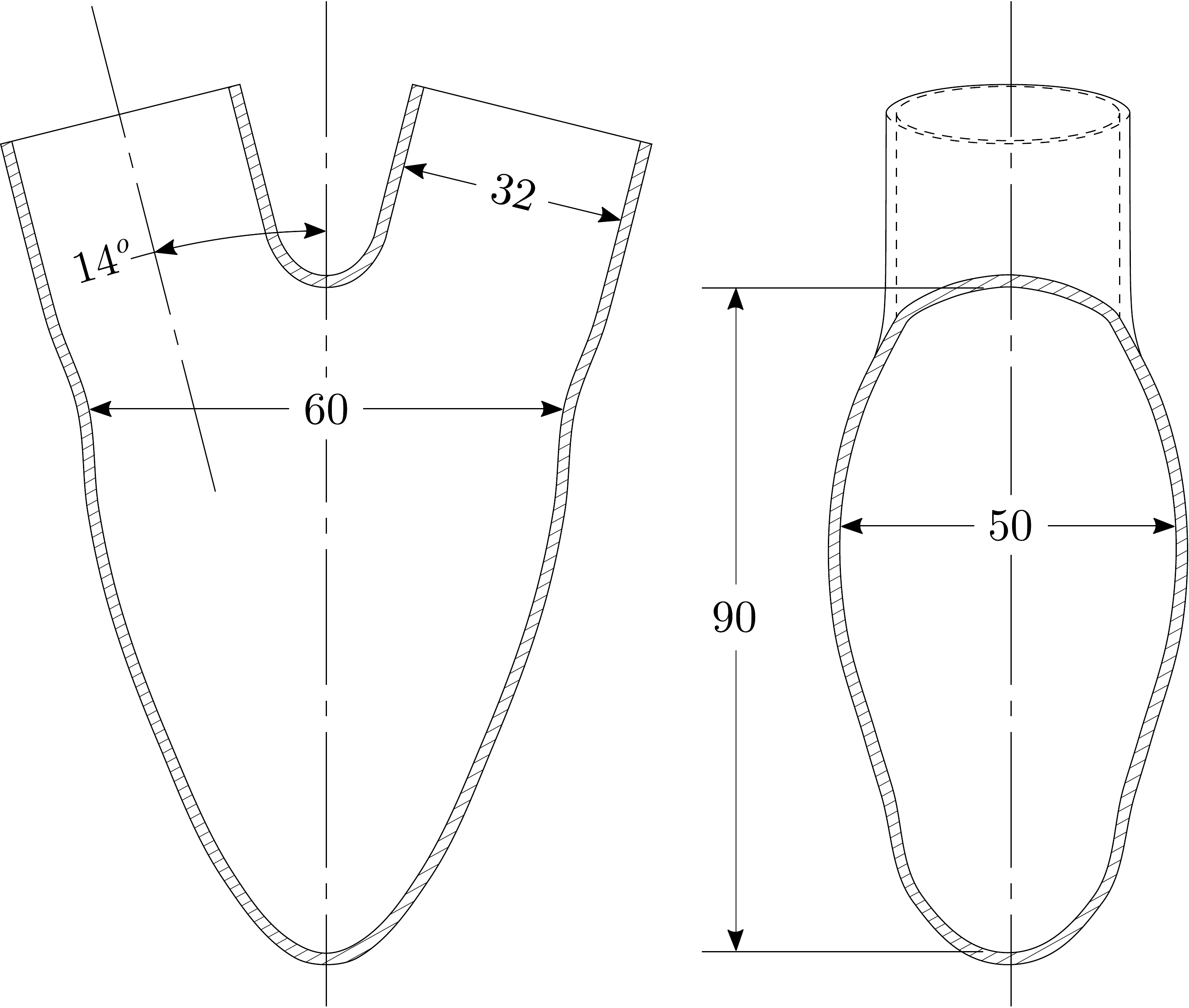}
		\caption{Symmetric left ventricular model used in this study.}
		\label{fig:LVdims}
		\end{figure}
		
	\section{\label{sec:Methods}Methodology}
	
		% Write here (remove lipsum line below).
		Regurgitant flow in the left ventricle was here studied using an in-house double activation left heart duplicator. The duplicator contains clear elastic silicone models of the left atrium (complete with its appendage and four pulmonary veins), left ventricle, and aorta (complete with sinuses of Valsalva). The choice of silicone (SILASTIC\textsuperscript{TM} RTV-$4234$-T$4$, The Dow Chemical Company, USA) is ideal for such an application, having low hardness ($40$ Shore A), easy moldability, forgiving elastic properties ($400$\% elongation at break, $27$-N/mm tear strength), and particle image velocimetry (PIV)-compatible optical properties when cured (optically transparent, refractive index of $1.41 \pm 0.01$). Although the simulator can accommodate a range of left ventricular geometries, the left ventricle used in this study was simplified and made symmetric while maintaining characteristic geometrical features in the middle to upper end of the size spectrum \cite{Lang06, Maceira06, Gibson14, Kou14}: end-diastolic volume of $155$ mL, base-to-apex length of $90$ mm, maximum width of $70$ mm, maximum depth of $50$ mm, and inflow-to-outflow tract angle of $14^{\circ}$ (see Fig.\ \ref{fig:LVdims}). In this way, the present study can focus on general qualitative flow features developing in the left ventricle without the introduction of patient-specific geometrical effects. The mitral valve ($23$ mm Perimount Magna Ease, Edwards Lifesciences, USA) is placed with its axis vertically at the base of the inflow tract and extends $15$ mm into the ventricle while the aortic valve ($25$ mm Perimount Theon RSR, Edwards Lifesciences, USA) is placed at the end of the left ventricular outflow tract. A mixture of $60$\% water and $40$\% glycerol by volume (refractive index of $1.39$) is used as the blood analog having a measured density ($1100$ kg/m$^3$) and dynamic viscosity ($4.2$ cP) comparable to those of blood at the working temperature of $23.1 \pm 0.2$ $^{\circ}$C.
		
		\begin{figure}[!h]
			\centering
			\includegraphics[width=0.75\linewidth]{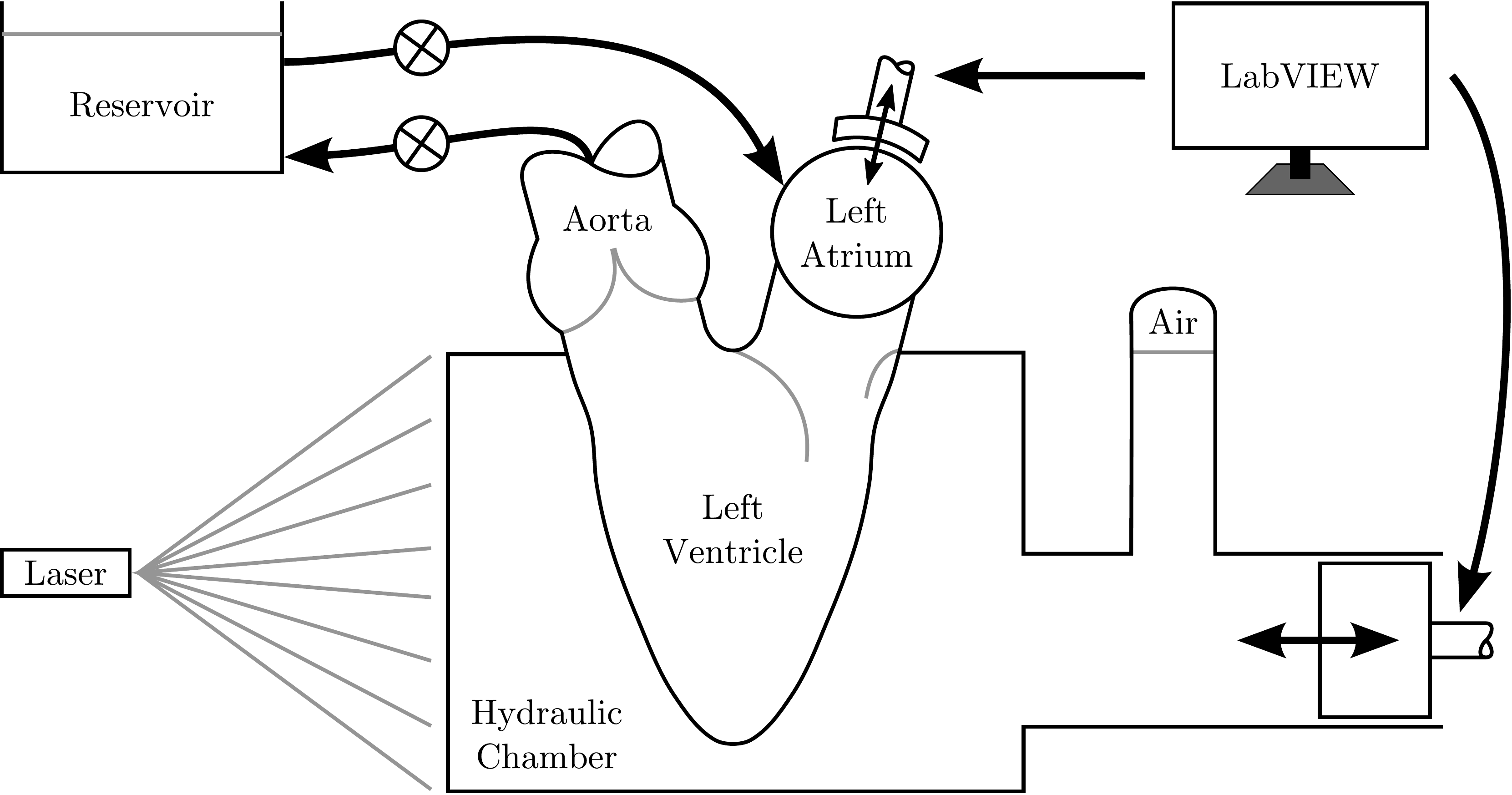}
		\caption{Schematic of the left heart duplicator. As the piston pulls back, the left ventricle fills through the mitral valve directly from the left atrium (and reservoir). Once the piston attains its maximal backward position, the left atrium is compressed to inject additional fluid into the ventricle (\textit{A} wave filling). This begins the piston's forward stroke, compressing the ventricle in the hydraulic chamber and ejecting fluid through the aortic valve up to the reservoir.}
		\label{fig:DupSchem}
		\end{figure}
		
		The functioning of the left heart duplicator works for the most part as many others do \cite{Mouret00, Kheradvar06, Okafor15}, with the added functionality of independent atrial and ventricular activation (see Fig.\ \ref{fig:DupSchem} for a schematic). The silicone left ventricle is encased in an acrylic hydraulic chamber, filled with the same working fluid to minimize optical distortion and fitted to a piston-cylinder assembly and a small column of air whose purpose is twofold, namely, to allow for some adjustment of the ventricular compliance and to accommodate the volume contributed from \textit{A} wave filling (atrial activation). The piston is controlled using an electromagnetically driven precision linear motor (LinMot, NTI AG, Switzerland) and a multifunction input-output device (PCI-$6281$, National Instruments, USA) through a custom graphical user interface built in LabVIEW (National Instruments, USA). Forward motion of the piston compresses the ventricle for ejection (systole) and backward motion expands the ventricle for filling (diastole). Atrial filling occurs passively from a reservoir. Notably, ventricular filling occurs in two distinct energetic phases, the \textit{E} wave, corresponding to relaxation of the ventricle through the action of the heart muscle, and the \textit{A} wave, corresponding to ejection of the atrium, providing an additional volumetric load to the ventricle prior to its ejection. In the present system, the piston produces only the \textit{E} wave of filling while the \textit{A} wave is achieved by physically compressing the atrium through a cam-follower mechanism controlled using a servomotor (Dynamixel RX-$24$F, Robotis, USA) and timed with the LabVIEW interface. This independent actuation of the \textit{E} and \textit{A} waves (referred to as double activation) is critical in simulating a disease like aortic regurgitation. In this disease, expansion of the ventricle due to relaxation of the heart muscle (the \textit{E} wave) results in the ventricle filling from both the mitral valve and the leaking aortic valve. This is then followed by ejection of an additional blood volume from the atrium into the ventricle (the \textit{A} wave) during which very little blood volume is contributed from regurgitation. Many existing heart duplicators produce both phases of ventricular filling hydraulically (using the piston), which would incorrectly result in both excessive and prolonged regurgitation; this effect is avoided entirely in our double activation system. The combination of the silicone aorta mold and the soft silicone tubing used throughout the system ($35$ Shore A hardness) provide enough compliance to the system to achieve healthy conditions while a peripheral resistance returning the aortic outflow to the reservoir allows for some additional tuning.
		
		\begin{figure}[!h]
			\centering
			\subfloat[\label{fig:HealthyFlow}]{%
				\includegraphics[width=0.45\linewidth]{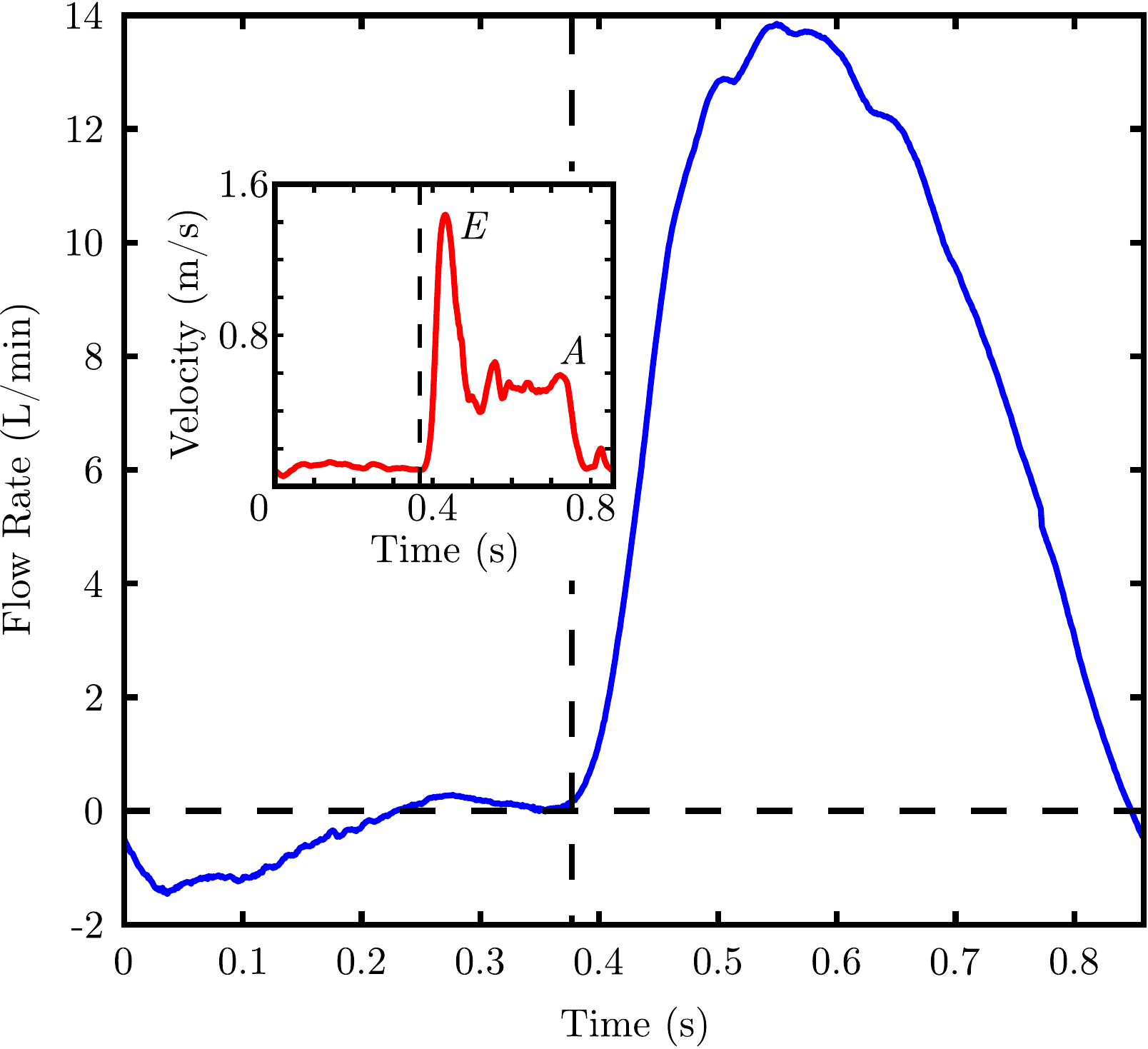}
			}\hfill
			\subfloat[\label{fig:HealthyPress}]{%
				\includegraphics[width=0.45\linewidth]{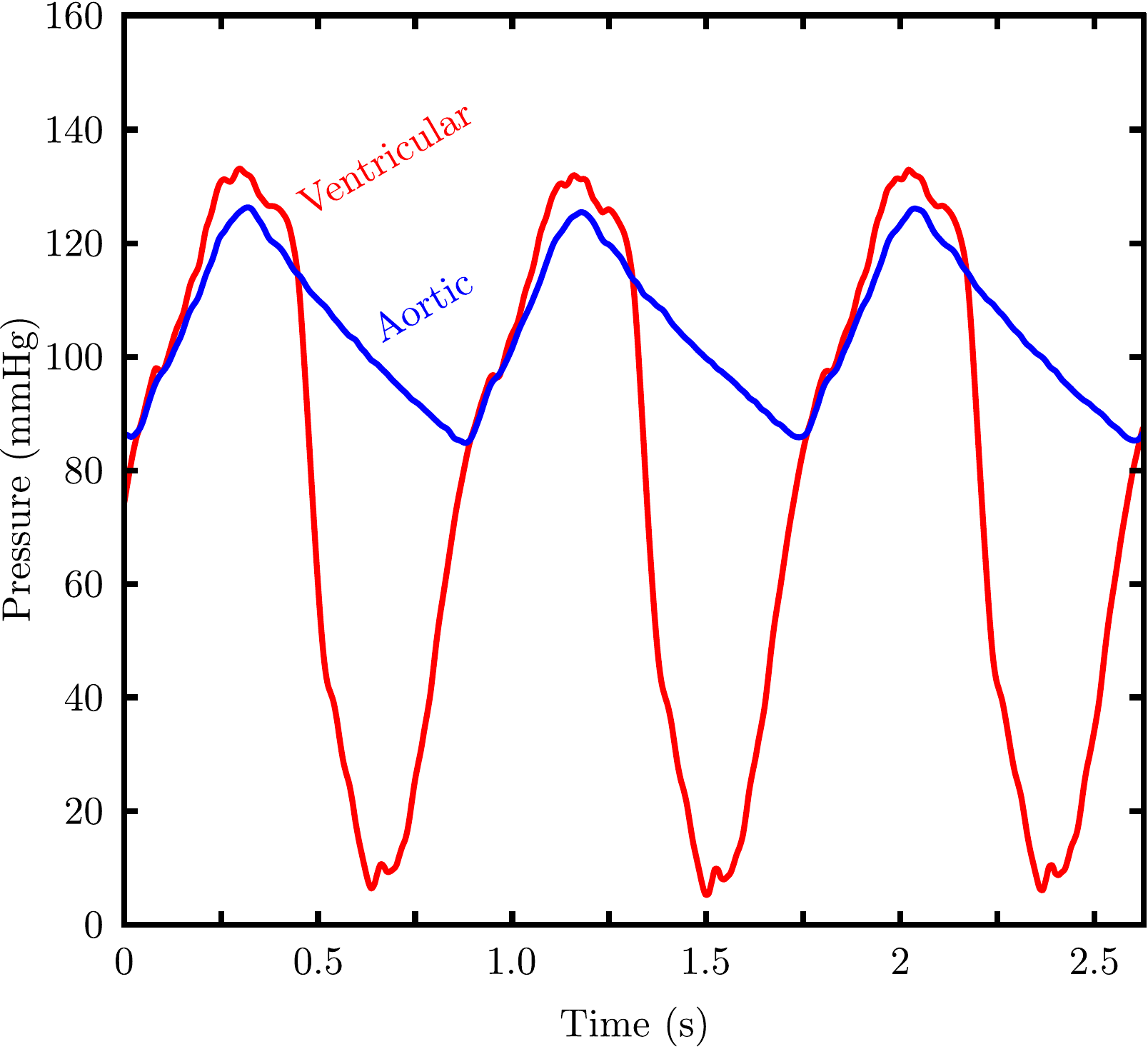}
			}
		\caption{(a) Mitral inflow waveform recorded just ahead of the reservoir generating a total cardiac output of $4.5$ L/min with distinct \textit{E} and \textit{A} waves of filling observed in the velocity field just below the mitral valve (inset). (b) Ventricular and aortic pressure waveforms falling in the expected healthy range. Note that the quantity $t^* = t/T$ is defined in Sec.\ \hyperref[sec:EulPatt]{\ref*{sec:ResDisc}.\ref*{sec:EulPatt}} and used thereafter, with $T = 0.857$ s being the cycle period, therefore $t = 0.857$ s in the above figures corresponds to $t^* = 1$.}
		\label{fig:HealthyConds}
		\end{figure}
		
		The duplicator reproduces healthy conditions rather well. At a heart rate of $70$ bpm and a stroke volume of $64$ mL, a cardiac output of $4.5$ L/min is recorded (the corresponding flow rate waveform is shown in Fig.\ \ref{fig:HealthyFlow}), the aortic and ventricular pressures fall in the healthy range (Fig.\ \ref{fig:HealthyPress}), and the mitral inflow shows distinct \textit{E} and \textit{A} waves (Fig.\ \ref{fig:HealthyFlow}, inset). It is to be noted that the flow rate was recorded just ahead of the reservoir, which is rather far upstream from the mitral valve, and so the \textit{A} wave of filling is not directly captured in Fig.\ \ref{fig:HealthyFlow}, while it can be seen by plotting the velocity just below the mitral valve in the inset of the figure. The corresponding intraventricular flow is also well reproduced, exhibiting the formation of the notorious vortex rolling up from the shear layer of the mitral jet during the \textit{E} wave (cf.\ Sec.\ \hyperref[sec:EulPatt]{\ref*{sec:ResDisc}.\ref*{sec:EulPatt}}). Regurgitation is introduced in the system by pulling apart $18$-gauge graduated rods ($\sim1$ mm in diameter) hooked under each aortic valve leaflet to create a central regurgitant orifice of known area; additional information regarding the simulator and regurgitation mechanism is available in the Supplemental Material of \citeA{DiLabbioKadem18}. In such a manner, we have simulated five cases of aortic regurgitation including one healthy scenario having regurgitant orifice areas (ROAs) of $0$ cm$^2$ (healthy), $0.10$ cm$^2$ (mild), $0.18$ cm$^2$ (moderate-$1$), $0.25$ cm$^2$ (moderate-$2$), $0.52$ cm$^2$ (severe-$1$), and $0.78$ cm$^2$ (severe-$2$); the severities were classified based on the AHA/ACC clinical guidelines in \citeA{Nishimura14}. These orifice areas represent $0$\%, $3.3$\%, $5.9$\%,
		$8.5$\%, $17.2$\%, and $26.1$\% of the fully open geometric aortic valve area ($3.00$ cm$^2$). As mentioned previously, in chronic aortic regurgitation, the left ventricle dilates to accommodate the regurgitant volume in an attempt to maintain a constant forward stroke volume. As such, the forward stroke volume was held constant at $64 \pm 4$ mL at a constant heart rate of $70$ bpm (cycle period $T = 0.857$ s). The mean mitral inflow Reynolds number among all cases is $1075 \pm 60$ and is computed as $\mathrm{Re} = 4{\rho}\dot{V}/{\pi}{\mu}d$, with $\dot{V}$ being the average flow rate. The associated Womersley number of the mitral inflow, defined as $\mathrm{Wo} = (d/2)\sqrt{2{\pi}{\rho}f/\mu}$ with $f$ being the cardiac frequency, is $15.9$. The peak aortic pressure was also held constant at $121 \pm 5$ mm Hg, which in cases of chronic regurgitation is known to either increase \cite{Bekeredjian05, Stout09} or remain relatively unchanged due to compensatory mechanisms such as the sympathetic nervous system \cite{Galbraith15}. The parameters of the experiment are summarized in Table \ref{tab:Simchars}.
		
		\begin{table}[!h]
			\begin{center}
				\begin{tabular}{llcll}
					\multicolumn{5}{p{\textwidth}}{\textbf{Table \ref*{tab:Simchars} -- Summary of Experimental Conditions.} For the regurgitant inflow, the orifice diameter was taken as $d = \sqrt{4A/\pi}$ for the calculation of the Reynolds ($\mathrm{Re}$) and Womersley ($\mathrm{Wo}$) numbers with $A$ being the regurgitant orifice area. The regurgitant fraction is the regurgitant volume divided by the stroke volume.} \\
					\hline\hline\\[-1.0em]
					\multicolumn{2}{l}{Working fluid} & & \multicolumn{2}{l}{Simulator} \\
					\hline\\[-1.0em]
					water-glycerol ratio & $60$:$40$ (by volume) & & cardiac output & $4.5 \pm 0.3$ L/min \\
					density $\rho$ &  $1100$ kg/m$^3$ & & cycle period $T$ & $0.857$ s ($70$ bpm) \\
					dynamic viscosity $\mu$ & $0.0042$ Pa s ($4.2$ cP) & & forward stroke volume & $64 \pm 4$ mL \\
					refractive index & $1.39$ & & mitral inflow mean $\mathrm{Re}$ & $1075 \pm 60$ \\
					temperature & $23.1 \pm 0.2$ $^{\circ}$C & & mitral inflow peak $\mathrm{Re}$ & $3500 \pm 500$\\
					& & & mitral inflow $\mathrm{Wo}$ & $15.9$ \\
					& & & nominal mitral valve diam & $23$ mm \\
					& & & nominal aortic valve diam & $25$ mm \\
					& & & peak aortic pressure & $121 \pm 5$ mm Hg \\
					\multicolumn{5}{l}{Regurgitation parameters} \\
					\hline\\[-1.0em]
					\multicolumn{2}{l}{regurgitant orifice areas} & & \multicolumn{2}{l}{$0$\%, $3.3$\%, $5.9$\%, $8.5$\%, $17.2$\%, $26.1$\% (of $3.00$ cm$^2$)} \\
					\multicolumn{2}{l}{diastolic aortic pressures} & & \multicolumn{2}{l}{$64$, $53$, $50$, $18$, $22$, $6$ mm Hg} \\
					\multicolumn{2}{l}{regurgitant fractions} & & \multicolumn{2}{l}{$0$, $0.11$, $0.34$, $0.40$, $0.47$, $0.52$} \\
					\multicolumn{2}{l}{regurgitant inflow peak $\mathrm{Re}$} & & \multicolumn{2}{l}{$0$, $9970$, $14\, 400$, $11\, 300$, $12\, 100$, $9460$} \\
					\multicolumn{2}{l}{regurgitant inflow $\mathrm{Wo}$} & & \multicolumn{2}{l}{$0$, $2.1$, $2.8$, $3.4$, $4.8$, $5.9$} \\
					\hline\hline
				\end{tabular}
			\end{center}
			\refstepcounter{table}\label{tab:Simchars}
		\end{table}
		
		The time-resolved velocity fields over one cycle for all cases were acquired using two-dimensional PIV in the plane bisecting the valve centers and ventricle apex (i.e., the view shown on the left in Fig.\ \ref{fig:LVdims} and in the schematics of Figs.\ \ref{fig:LVschem} and \ref{fig:DupSchem}). This plane is similar to the parasternal long axis and apical three-chamber views commonly used in clinical practice to assess aortic regurgitation via ultrasound \cite{LancellottiTribouilloy13}. Additionally, while the flow is certainly inherently three dimensional, this plane represents a plane of symmetry for the ventricular geometry used here and as such ought to represent an approximately two-dimensional flow. The flow was seeded with $50$-$\upmu$m polyamide particles (Dantec Dynamics, Denmark) with a density of $1030$ kg/m$^3$. The particles were illuminated using a double-pulsed Nd:YLF laser (LDY$301$, Litron Lasers, England) emitting $100$-ns pulses of coherent light of wavelength $527$ nm at a repetition rate of $0.2$-$20$ kHz. The beam was guided from the source to the hydraulic chamber via an articulated guiding arm (LaVision GmbH, Germany) where it was passed through a cylindrical lens to form a $1$-mm-thick light sheet. The left ventricular flow domain was captured using a complementary metal-oxide semiconductor (CMOS) camera (Phantom v$9.1$, Vision Research Inc., USA), recording double-frame images $700$ $\upmu$s apart at $400$ Hz at the full camera resolution of $1632 \times 1200$ pixels. In the acquired images, the ventricle measured $76$ mm in height and $65$ mm at maximum width (the ventricle base width just below the mitral valve leaflet tips). The velocity fields were computed using DaVis $8.2$ (LaVision GmbH, Germany) by processing the double-frame images using the multipass cross-correlation algorithm with decreasing interrogation window size and a standard fast Fourier transform for the correlation function. The interrogation window size for the initial passes was $64 \times 64$ pixels with $50$\% overlap and uniform weighting, which was reduced to $16 \times 16$ with $50$\% overlap for the final passes with a round Gaussian weighting function, corresponding to a final spatial resolution of $0.52 \times 0.52$ mm$^2$. The simulator was left to operate for $50$ cardiac cycles prior to any acquisition to ensure minimal cycle-to-cycle variation. Additionally, the time-resolved acquisitions were performed ten times for each case and, given the repeatability of the experiment, ensemble averaged to filter out the remaining minor velocity fluctuations. The total uncertainty in the velocity fields resulting from the PIV system setup and the subsequent measurements were estimated to be below $5$\% with respect to the maximum pointwise velocities in the healthy ($1.70$ m/s) and most severe ($1.66$ m/s) cases based on the major sources of uncertainty described in \citeA{Raffel07} and \citeA{Adrian11}.
		
	\section{\label{sec:ResDisc}Results \& Discussion}
	
		\subsection{\label{sec:EulPatt}Eulerian flow patterns arising from aortic regurgitation}
		
			With aortic regurgitation, the emergence and interaction of two pulsatile inflowing jets prompts rather interesting dynamics. In this scenario, the mitral jet contributes a constant volume to filling while the volume contributed by the regurgitant jet increases with severity. In our previous work \cite{Raymondet16, DiLabbio16, BenAssa17, DiLabbioKadem18}, we have demonstrated that the left ventricular filling vortex may gradually reverse direction as a result of increasing regurgitation severity, which was captured by both the velocity field and a gradual change in sign of the circulation, and that the dissipation of energy in the left ventricle due to viscosity consequently increases monotonically. While in this work we strictly aim to understand the material transport behavior in the left ventricle under the conditions of aortic regurgitation, the physics of which remain to be investigated, in this section we offer a brief overview of the general Eulerian flow patterns observed in our previous work \cite{DiLabbioKadem18}. In order to distinguish the phases of the cardiac cycle in a simple manner, we define a nondimensional time $t^* = t/T$ such that $t^* = 0$ marks the beginning of the ejection phase (end of the filling phase for the previous cycle), $t^* = 0.438$ marks the end of the ejection phase (beginning of the filling phase), and $t^* = 1$ marks the end of the cycle (end of the filling phase, beginning of the ejection phase for the next cycle). For later use, we will additionally define two constants, namely, $\alpha = 0.562$, representing the fraction of the cardiac cycle corresponding to the filling phase, and $\beta = 0.438$, representing the fraction corresponding to the ejection phase.
			
\begin{figure}[!h]
				\centering
				\includegraphics[width=0.80\linewidth]{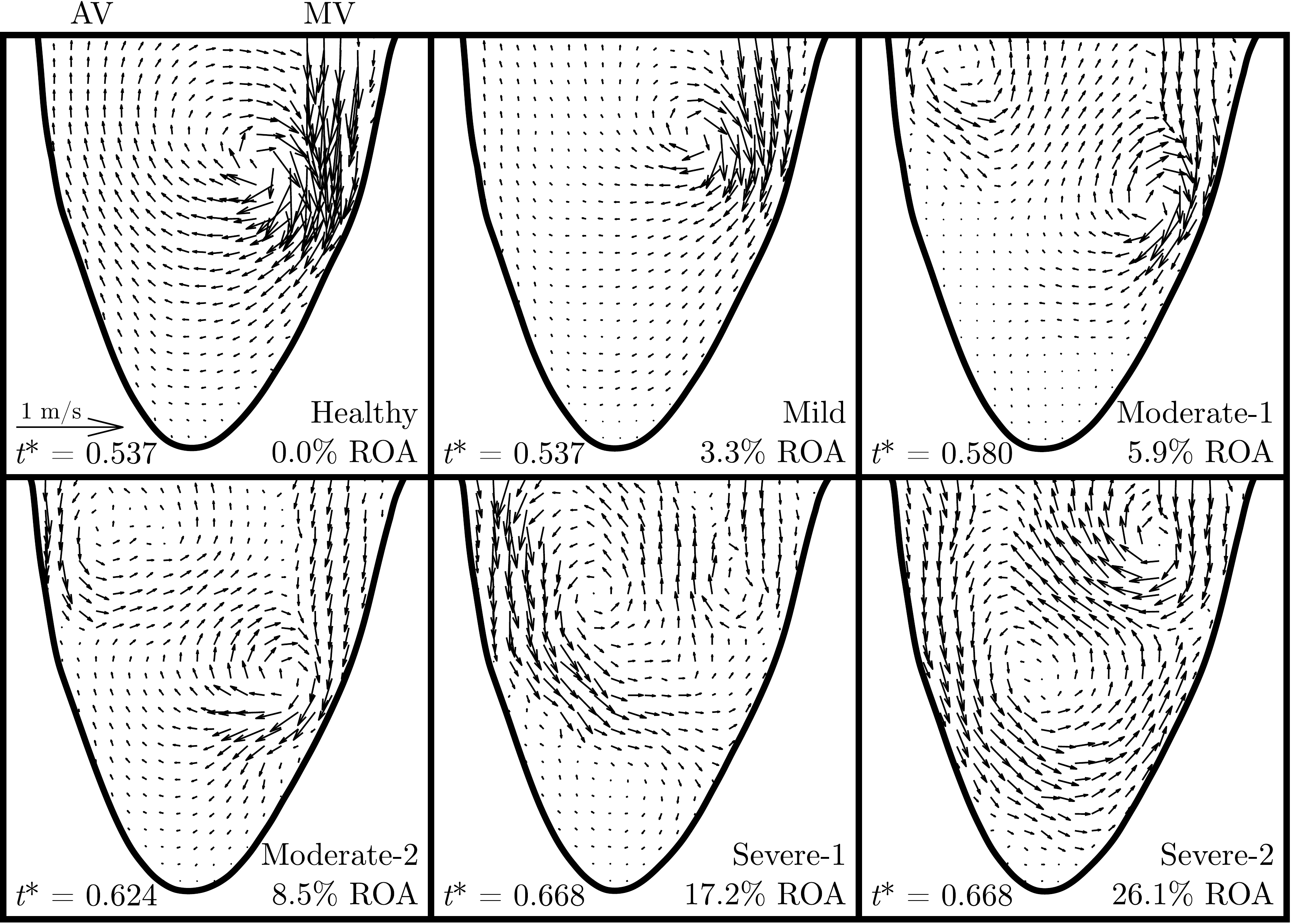}
			\caption{Intraventricular flow patterns at selected instants during the left ventricle's filling phase for all simulated cases of aortic regurgitation. Note the progression from a dominant mitral jet-driven vortex, imparting a clockwise swirl to the ventricular volume, to a dominant regurgitant jet-driven vortex, imparting a counterclockwise swirl.}
			\label{fig:EulFlow}
			\end{figure}

			Healthy left ventricular filling during the \textit{E} wave is characterized by a jet emanating from the mitral valve, generating a vortex ring with one end quickly dissipating against the ventricular wall and the other end persisting as a clockwise vortical structure throughout the duration of the cardiac cycle, imparting a clockwise swirl to the entire ventricular blood volume. The vortex is marked by the skewed velocity profile typical of vortex rings, with near solid-body rotation in the core and the jet end of the vortex having higher peak velocity than the opposite end. This vortex sets up in the ventricle's center and persists until ejection, providing an optimal kinetic-energy-conserving flow path from inflow to outflow. The early progression of this vortex can be seen in Fig.\ \ref{fig:EulFlow} (top left), along with the intraventricular flow for all regurgitant cases at selected instants during the filling phase. In the case of mild aortic regurgitation ($\mathrm{ROA}$ equal to $3.3$\%), both the regurgitant orifice area and regurgitant volume are rather small. As a result, the regurgitant volume is simply entrained by the mitral inflow and the flow somewhat resembles that of a healthy ventricle with less forceful filling, as shown by the slower downstream progression of the filling vortex in Fig.\ \ref{fig:EulFlow} (top center), plotted at the same instant as in the healthy scenario for comparison. As chronic aortic regurgitation enters the moderate stages, a significant difference can be observed in the flow field. The regurgitant jet transports enough volume to generate its own vortex rotating counter to that driven by the mitral inflow. The two vortices effectively compete for space within the ventricular volume, with the regurgitant jet-driven vortex gradually dominating as the severity worsens (see the remaining four panels of Fig.\ \ref{fig:EulFlow}). It can be observed that the resulting dynamics is strongly dependent not only on the regurgitant volume and orifice area, but also on the timing of the onset of filling from each jet. This was evidenced between the moderate-$1$ ($\mathrm{ROA}$ equal to $5.9$\%) and moderate-$2$ ($\mathrm{ROA}$ equal to $8.5$\%) cases, where in the moderate-$1$ case the regurgitant jet emanates after the mitral inflow, visibly nudging the mitral jet-driven vortex towards the ventricular wall on the mitral side, which can be seen to some extent in Fig.\ \ref{fig:EulFlow} (top right). The timing and speed of the regurgitant jet permitted the regurgitant volume to penetrate deeper into the ventricle, while remaining confined to the wall on the aortic side (this feature will be evidenced in Secs.\ \hyperref[sec:BloodTrans]{\ref*{sec:ResDisc}.\ref*{sec:BloodTrans}} and \hyperref[sec:BloodTrans]{\ref*{sec:ResDisc}.\ref*{sec:LCSs}}). The difference in timing between the two jets in the moderate-$2$ case however, where the regurgitant and mitral jets emanate almost at the same time instant, does not permit the same penetration depth of the regurgitant volume. This behavior suggests the existence of a critical condition, dependent upon jet timing and strength, separating distinct intraventricular dynamics. With severe aortic regurgitation, the regurgitant jet clearly emanates before the mitral jet and consequently sets up the more dominant counterrotating vortex, imparting a completely reversed swirl to the majority of the ventricular volume during the early filling phase. In all cases, with the majority of the regurgitation occurring during the \textit{E} wave of filling, the mitral jet continues to oppose the effects of the regurgitation throughout the remainder of the filling phase, though never enough to establish the pattern observed in the healthy ventricle, namely, to impart a clockwise swirl to the entire ventricular fluid volume.
			
			\begin{figure}[!t]
				\centering
				\includegraphics[width=0.80\linewidth]{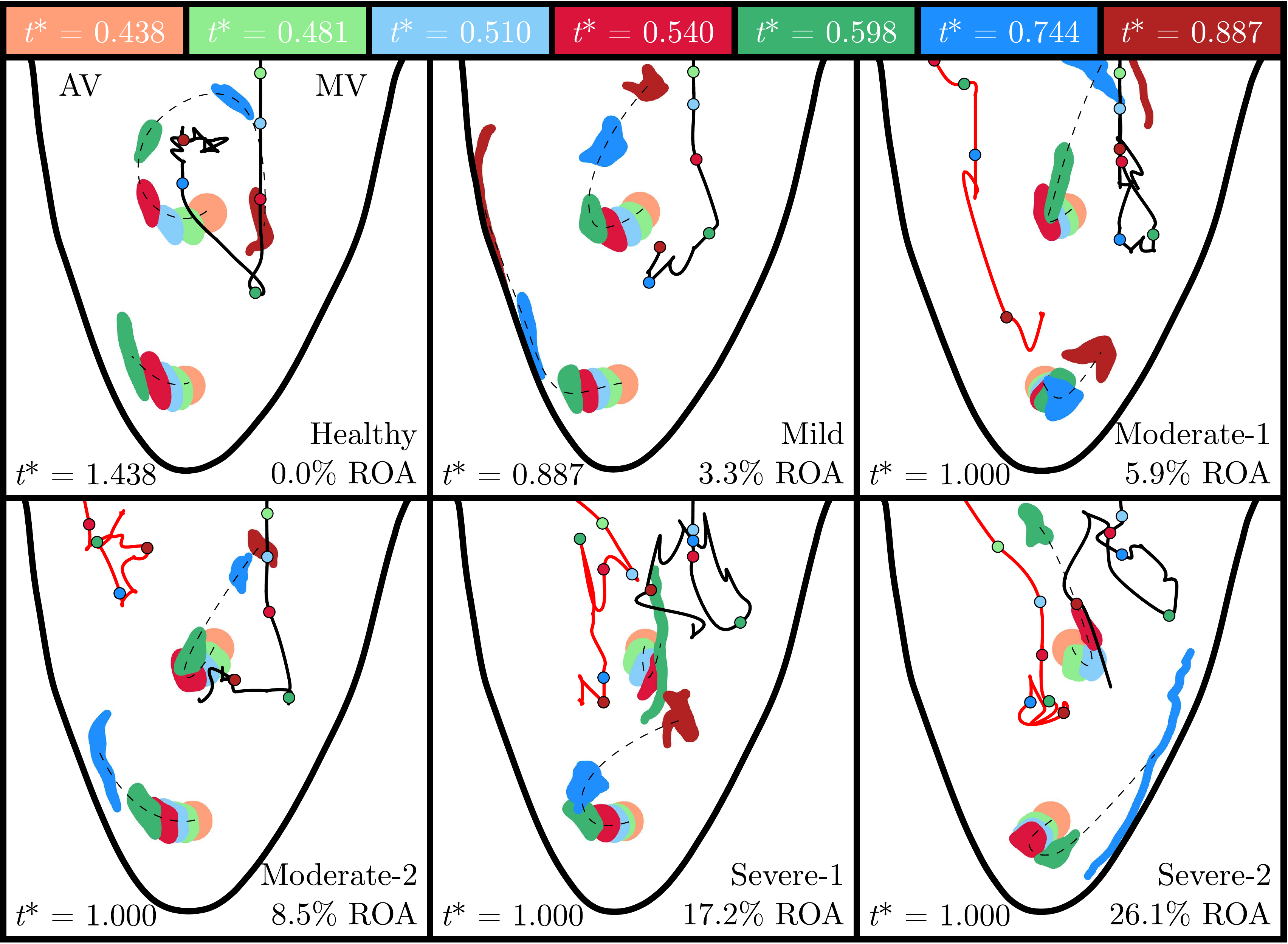}
			\caption{Path followed by the vortex cores generated during the \textit{E} wave of filling, from the start to the end of the filling phase (i.e., from $t^* = 0.438$ to $t^* = 1$), for all cases simulated in this study. The mitral vortex core path is traced in black, while that of the regurgitant vortex core is traced in red. The advection of two circular blobs, one in the ventricle center and one in the apical region, is also illustrated, color coded according to the vortex core positions at specific time instants. The dashed lines show approximately the motion of the blob centroids as a visual aid. Note that the mitral vortex in the healthy ventricle is tracked for one complete cycle (i.e., from $t^* = 0.438$ to $t^* = 1.438$), while that in the case of mild regurgitation is tracked until $t^* = 0.887$, where the vortex is sheared out.}
			\label{fig:VortexTraces}
			\end{figure}
			
			As has been suggested in several previous works (cf.\ \citeA{PedrizzettiDomenichini15}), the healthy swirling flow in the left ventricle is believed to optimally align material for ejection. With two jet-driven counterrotating vortices interacting in a confined geometry, their resulting impact on material transport and organization poses a rather intriguing fluid dynamics problem. It would therefore be of use to observe and track the relative motion between the two vortex cores during the filling phase in order to gain some idea as to how these organizing centers dictate transport. The vortex core positions were identified using peaks of the $\Gamma_2$ function described in \citeA{Michard97} and \citeA{Graftieaux01} and then followed from the moment the cores were identifiable after the start of filling ($t^* = 0.438$) until the end of the filling phase ($t^* = 1$) or until the cores were no longer traceable due to excessive shearing or vortex breakdown. This function effectively provides a Galilean invariant measure of the presence of a local center of fluid rotation in a series of interrogation regions, admitting values in the range $[-1,1]$ with negative values signifying a center for clockwise rotation and positive values for counterclockwise rotation (alternatively, it is also common to use the $\Gamma_2$ function in absolute value). The corresponding vortex core paths are shown in Fig.\ \ref{fig:VortexTraces} along with the advection of two circular blobs, one in the center of the ventricular geometry and one in the apical region, colored according to the vortex core positions at specific time instants. The reader is implored to use the color order to navigate themselves along the vortex path in time, particularly due to their convoluted nature with the presence of aortic regurgitation. For the healthy ventricle (Fig.\ \ref{fig:VortexTraces}, top left), the \textit{E} wave vortex core is tracked for one complete cycle (i.e., from $t^* = 0.438$ to $t^* = 1.438$). The core is first seen to propagate almost directly downward, attaining a peak velocity of $0.58$ m/s and forming a channel with the wall to guide the incoming mitral flow down to the ventricle apex. With the downward motion of the core, the circular blob in the apical region is swept and stretched toward the ventricle wall on the aortic side. The core then changes course toward the ventricle center, arriving at approximately $t^* = 0.796$ and hovering in position at more or less the same height for the remainder of the cycle until it is ultimately forced to merge with the next \textit{E} wave vortex. The overall motion of the core from the start to the end of the filling phase coordinates an elegant swirl of the circular blob in the ventricle center, ultimately aligning the material with the inflow by $t^* = 0.887$ and allowing the centrally located vortex to guide this material toward the left ventricle outflow tract. With mild regurgitation (Fig.\ \ref{fig:VortexTraces}, top center), a concentrated vortex is still seen to form at the start of filling, however propagating downstream more slowly (evidenced by the progression of the color markers when comparing the healthy and mild cases in Fig.\ \ref{fig:VortexTraces}) and attaining an overall shorter penetration depth. The vortex core follows the same vertical path as in a healthy ventricle at first (until $t^* = 0.540$), attaining a peak velocity of $0.44$ m/s prior to being deflected toward the ventricle wall on the mitral side, but is ultimately unable to set up in the ventricle's center, effectively being destroyed around $t^* = 0.887$. The deflection of the vortex core path around $t^* = 0.540$ coincides with the onset of the regurgitant jet, although the jet itself cannot be seen in the PIV measurements due to its path taken in the $14$-mm region above the measurement domain arising from the rightward velocity imparted to it by the clockwise mitral vortex; the regurgitant volume is subsequently entrained by the mitral inflow. Although the vortex core is effectively sheared out before the end of the filling phase, the course taken by the vortex core still permits the apical blob to be organized for ejection. The movement of the central blob, however, does not follow the smooth swirl observed in the healthy left ventricle; rather it is deflected far upstream, toward the inlet of the mitral valve where it will be considerably sheared by the mitral jet for the remaining duration of the filling phase. The remaining four panels of Fig.\ \ref{fig:VortexTraces} show the vortex core paths for the moderate and severe cases, with both the regurgitant and mitral jet-driven vortices being tracked. For the moderate-$1$ case ($\mathrm{ROA}$ equal to $5.9$\%), the onset of the regurgitant jet is clearly seen to coincide with the deflection of the mitral vortex core toward the ventricle wall at $t^* \approx 0.540$. The ability of the regurgitant vortex core to penetrate deeper into the ventricle is also observable in the figure when compared to the other regurgitant cases. While the central material blob exhibits behavior similar to that in the mild case (note the aforementioned shearing occurring due to the mitral jet for $t^* = 0.744$ and $t^* = 0.887$), the apical blob is driven away from the ventricle outflow tract due to the regurgitant jet and the weak counterclockwise vortex setting up in the apex. In the bottom row of Fig.\ \ref{fig:VortexTraces}, the mitral jet-driven vortex core is seen to propagate less downstream during the \textit{E} wave (approximately from $t^* = 0.438$ to $t^* = 0.510$) with regurgitation severity, a feature which appears to be generally related to impeded ventricular filling, observed both \textit{in vivo} in the case of diastolic dysfunction \cite{Charonko13} and \textit{in silico} in the case of an infarcted ventricle \cite{Badas17}. On the contrary, the regurgitant jet-driven vortex core propagates further downstream, ultimately setting up above the ventricle apex for severe regurgitation (also seen in the bottom row of Fig.\ \ref{fig:EulFlow}). With the regurgitant jet in the moderate-$2$ case ($\mathrm{ROA}$ equal to $8.5$\%) emanating earlier than in the moderate-$1$ case, although still nudging the mitral vortex toward the ventricle wall, it is unable to benefit from the same effect and penetrates less deep into the ventricle, setting up a counterclockwise vortex residing in the vicinity of the aortic valve and impeding the upward motion of the mitral vortex. As a consequence, the circular blob in the apical region is directed by the clockwise swirl of the mitral vortex toward the outflow tract as in the healthy and mild cases. With severe regurgitation ($\mathrm{ROA}$ equal to $17.2$\% and $26.1$\%), it is the onset of the mitral jet which nudges the regurgitant vortex toward the ventricle wall on the aortic side ($t^* \approx 0.510$ in both cases). With the regurgitant vortex establishing itself above the apex, the circular blob in the apical region is again directed away from the outflow tract as in the moderate-$1$ case, with a slight difference in dynamics associated with the strength and timing of the regurgitant jet. The central circular blobs in the severe cases are subject to a strong upward and leftward motion from combined rotation of the mitral and regurgitant vortices and therefore are interestingly directed toward the outflow tract.
			
		\subsection{\label{sec:BloodTrans}Blood transport characteristics associated with aortic regurgitation}
		
\begin{figure}[!t]
				\centering
				\includegraphics[width=1.00\linewidth]{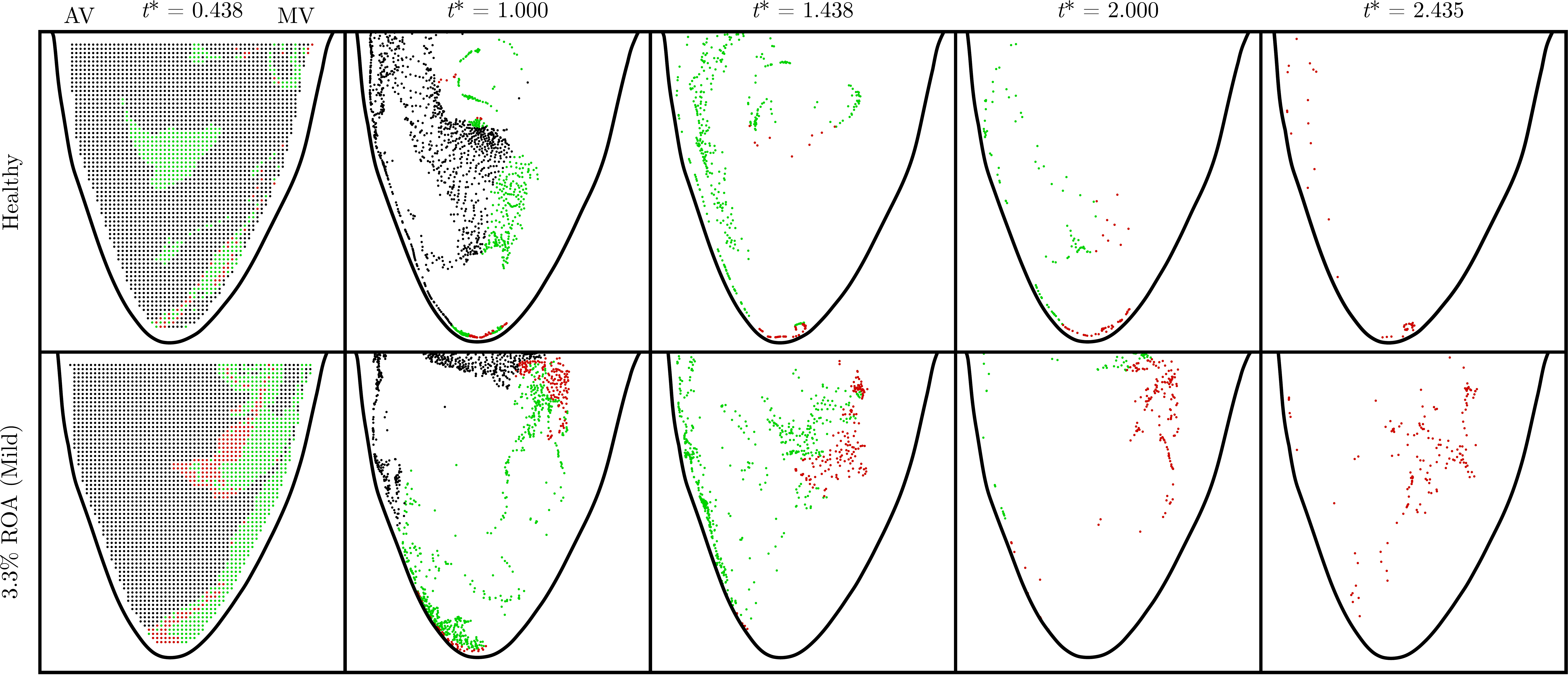}
			\caption{Advection of virtual particles from the beginning of the filling phase ($t^* = 0.438$) over two cycles for a healthy left ventricle (top row) and for one with mild aortic regurgitation having a regurgitant orifice area $3.3$\% of the fully open geometric aortic valve area (bottom row). Particles are colored black if they are ejected in the first beat, green if ejected in the second beat, and red if the particles remain after two beats. The first, third, and fifth columns represent the beginning of the filling phase (end of ejection) for consecutive cycles, while the second and fourth columns represent the beginning of the ejection phase (end of filling) for the respective cycles.}
			\label{fig:AdvNormMild}
			\end{figure}

			The tendency of the mitral vortex to set up in the ventricle's center appears to have a significant influence on proper material alignment of blood already present in the ventricle for ejection. To illustrate this influence, the flows in the healthy and mild ($\mathrm{ROA}$ equal to $3.3$\%) cases were seeded with $\sim2000$ virtual particles at the start of the filling phase ($t^* = 0.438$) and advected over two cycles. For all results reported in this section, the fourth-order Runge-Kutta scheme was used to advect virtual particles in time. The virtual particles were treated as material elements of the fluid itself, therefore any inertial effects associated with true blood particles were not considered in the calculations. When multiple cycles are involved, the data set for a single cycle was simply appended to itself as many times as needed. As an additional word of caution, it should be noted that no distinction is made during advection between particles fleeing the flow domain and particles truly being ejected. Figure \ref{fig:AdvNormMild} shows the resulting particle advection patterns associated with the healthy (top row) and mild (bottom row) cases. The first column of Fig.\ \ref{fig:AdvNormMild} marks all the particle positions at the start of advection, colored black if they are ejected in the first beat, green if ejected in the second beat, and red if ejected after two beats. By the end of the filling phase for the first cycle (Fig.\ \ref{fig:AdvNormMild}, second column), it is clear that the healthy left ventricle performs very well at aligning material for ejection, organizing the majority of the initial virtual particles (the black particles) into a columnar region under the outflow tract, which are then easily ejected. Additionally, a portion of material that will only be ejected in the subsequent beat (the green particles) has also been well aligned for ejection, trailing just behind the black particles, however the aortic valve closes just ahead of them and they are therefore not ejected. By the end of the ejection phase for the first cycle (Fig.\ \ref{fig:AdvNormMild}, third column), $15.8$\% of the initial particles remain, most of which are already aligned for ejection and are left to be either pushed toward the outflow tract or swirl around the vortex core. This pattern then repeats in the subsequent beat, where now green particles are organized into a column for ejection by the end of the filling phase (Fig.\ \ref{fig:AdvNormMild}, fourth column), and very few of the initial particles remain ($2.0$\%) at the end of the subsequent ejection phase (Fig.\ \ref{fig:AdvNormMild}, fifth column). With slight regurgitation, more particles reside in the ventricle over longer periods marked by more green and red regions in Fig.\ \ref{fig:AdvNormMild} (first column). The absence of a vortex core in the ventricle center causes an overall inefficiency in organizing particles for ejection. For instance, at the end of the filling phase for the first cycle (Fig.\ \ref{fig:AdvNormMild}, second column), very few particles are organized into a column under the outflow tract, and $37.8$\% of the initial particles remain after ejection (Fig.\ \ref{fig:AdvNormMild}, third column). This feature repeats in the subsequent cycle with $9.9$\% of the initial particles remaining at the end of ejection (Fig.\ \ref{fig:AdvNormMild}, fifth column). The reader is referred to the Supplemental Material for a dynamic view of particle advection behavior for all cases over two cardiac cycles. This elementary demonstration advocates the discussion of blood stasis in the left ventricle in the case of aortic regurgitation. In general, with blood consistently regurgitating back into the ventricle, it is natural to suspect that blood particles ought to reside in the ventricle for longer periods. Furthermore, the flow patterns observed for aortic regurgitation are considerably complex, lacking the simple coherent pattern offered by the mitral vortex in the healthy flow scenario, and therefore will characteristically delay blood from being ejected overall. This same question of the importance of the healthy left ventricular flow pattern on particle residence time may also have significant implications in the design of left ventricular assist devices, the flow patterns of which considerably deviate from those of a natural heart \cite{Long14}.
			
\begin{figure}[!t]
				\centering
				\subfloat[\label{fig:EjectionMapa}]{%
					\includegraphics[width=0.80\linewidth]{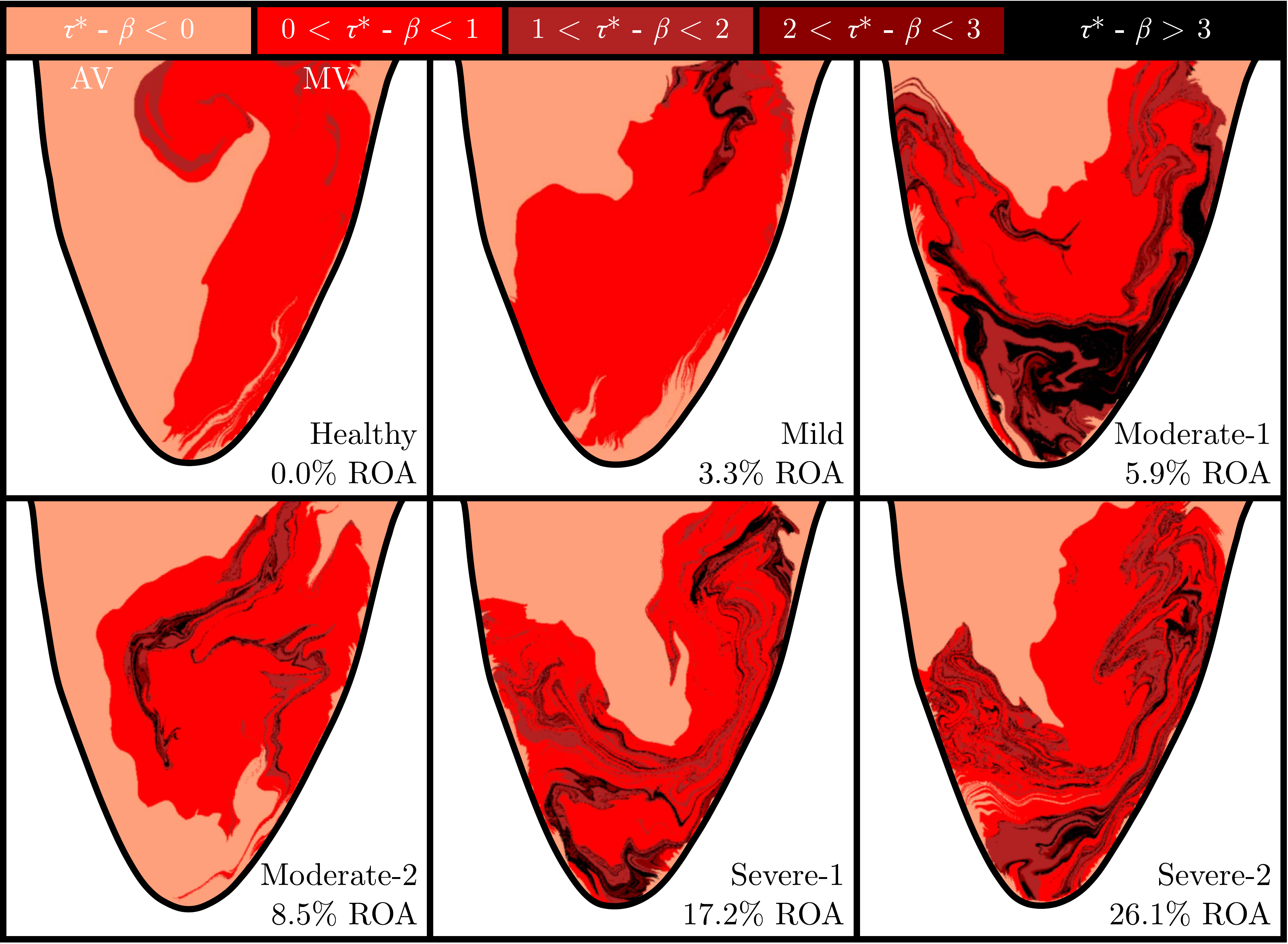}
				}\\
				\subfloat[\label{fig:EjectionMapb}]{%
					\includegraphics[width=0.98\linewidth]{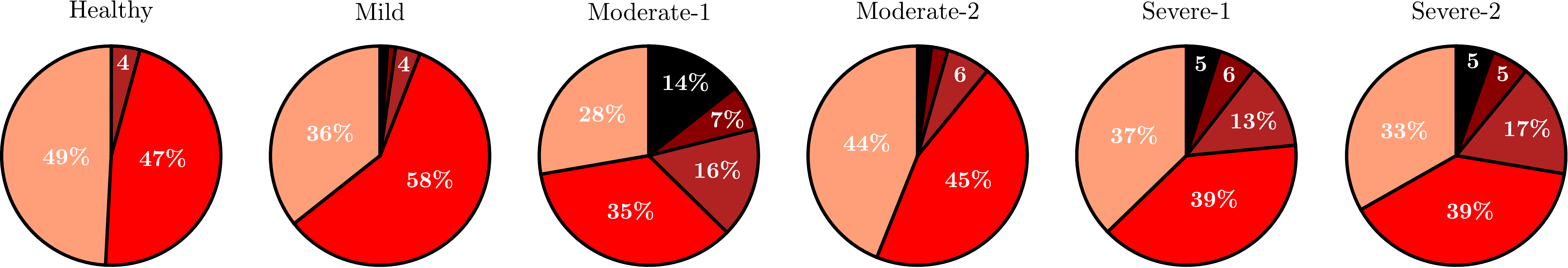}
				}
			\caption{(a) Normalized particle residence time ($\tau^*$) in the left ventricle at the start of the ejection phase ($t^* = 0$) colored by time until ejection in beats; $\beta = 0.438$ is the duration of the ejection phase normalized by the cycle period ($T = 0.857$ s). (b) Percentage of ventricle area corresponding to the colored regions in (a).}
			\label{fig:EjectionMap}
			\end{figure}

			Blood stasis in the left ventricle could promote rather serious adverse effects. The longer blood particles reside in the ventricle, the more prone activated platelets are to agglomerate. While the left ventricle itself is well known to avert the formation of mural thrombi, likely due to the highly dynamic nature of the intraventricular flow, platelets that have agglomerated in stagnant regions of the left ventricle may still be ejected and ultimately clot arteries, which in the worst case may result in ischemic stroke. With one-third of all ischemic strokes occurring in the USA being cryptogenic (of unknown cause) \cite{Yaghi14}, it is of paramount importance to identify possible causes and conditions. The possibility of platelet agglomeration (thrombus formation) in aortic regurgitation must be entertained, with the study of \citeA{Petty00}, for example, estimating a $7.5$\% incidence in cerebrovascular events in patients with aortic regurgitation after five years, the incidence of which increases in the presence of comorbidities (such as aortic stenosis). By observing the residence time of particles in the left ventricle, stasis of blood could be better understood in the case of aortic regurgitation. A map of Lagrangian particle residence time in the left ventricle, here denoted by $\tau$, can be computed by densely seeding the left ventricle with virtual particles ($\sim200\, 000$) at a selected time instant and advecting these particles until they flee the flow domain. As some particles may only be ejected after a large number of subsequent beats, advection of any given particle was terminated when its residence time surpassed four heart beats. Figure \ref{fig:EjectionMapa} displays the particle residence time at the start of the ejection phase ($t^* = 0$) with areas being colored by time until ejection in beats. The corresponding percentage of each colored region with respect to the total ventricle area is given in Fig.\ \ref{fig:EjectionMapb}. The aforementioned ability of the healthy ventricle to organize particles for ejection is accentuated, with $49.2$\% of the ventricular area being clearly organized and ejected in the first beat and $46.6$\% being ejected in the second beat, consistent with the findings of previous \textit{in vivo} studies \cite{Bolger07, Eriksson13}. There is a very small degree of blood stasis in a healthy left ventricle, a feature that is very well captured by the present \textit{in vitro} model, with only $4.2$\% of the ventricular area being ejected in subsequent beats. With mild regurgitation ($\mathrm{ROA}$ equal to $3.3$\%), as discussed previously, the mitral vortex persists almost until the end of the filling period and while it is unable to set up in the ventricle's center, some organization is still achieved with $35.7$\% of the ventricular area being ejected in the first beat and $58.4$\% being ejected in the second beat, leaving $5.9$\% to be ejected in subsequent beats. Notably, while the healthy left ventricle possessed virtually no regions that were ejected in more than three beats (less than $0.03$\%), even mild regurgitation is enough to increase stasis duration of some fluid elements by two or more additional beats ($2.3$\% ejected in over three beats). The moderate-$1$ case ($\mathrm{ROA}$ equal to $5.9$\%) presents the most striking and most interesting case. As the regurgitant jet managed to penetrate deeper into the ventricle while being confined to the wall on the aortic side, the weak regurgitant vortex sets up just above the ventricle apex and is further strengthened by the rotation of the mitral vortex. Although we have previously shown that the total energy dissipated per cycle increases almost linearly with regurgitation severity in this \textit{in vitro} model \cite{DiLabbioKadem18}, the developing flow patterns resulted in blood stasis being considerably worse in the moderate-$1$ case compared to all other cases, having $37.3$\% of the ventricular area being ejected in more than two beats. These fractions are in fact insensitive to the number of virtual particles advected, provided they are uniformly distributed within the left ventricle, with an observed percent difference of less than $5$\% between advecting $3000$ particles and $850\, 000$ particles and even less when rounding to two significant digits. The choice to advect a larger number of virtual particles ($200\, 000$) is however not arbitrary as enough must be advected to resolve the detail in the particle residence time map in Fig.\ \ref{fig:EjectionMapa} and nonetheless the fractions are more accurate, differing by less than $2$\% compared to advecting $850\, 000$ particles. The same is true of the results reported in the remainder of this section. It is to be noted that although the forward stroke volume was held constant for all simulated cases, the ejected areas of the ventricle may be expected to increase overall due to dilation of the ventricle to maintain the constant volume contributed from the mitral inflow and to accommodate the additional volume contributed from the regurgitant inflow. Expansion of the ventricle occurred largely in the out-of-plane direction however, and given that this is a two-dimensional slice of an inherently three-dimensional flow, this mismatch further stresses the importance of investigating the intraventricular flow in three dimensions. Nonetheless, with this in mind, it may be of interest to obtain the mean value of the particle residence time in beats, calculated at the start of ejection and limited to four beats, to obtain a single parameter expressing a tendency toward blood stasis, here denoted by $\mathrm{PRT}_2^4$ and defined by (\ref{eq:PRT24}) below. The variable $\beta$ again denotes the fraction of the cardiac cycle corresponding to the ejection phase and $\tau^* = \tau/T$. The value of $\mathrm{PRT}_2^4$ is reduced by setting the particle residence time to zero for particles ejected within two beats. This is done to take into account the large fraction of material volume ejected in the first two cycles for a healthy left ventricle; in this way values of $\mathrm{PRT}_2^4$ will be especially low for healthy patients. The values of $\mathrm{PRT}_2^4$ are $0.04$, $0.06$, $0.37$, $0.11$, $0.23$, and $0.28$, corresponding, respectively, to the cases with a $\mathrm{ROA}$ equal to $0$\%, $3.3$\%, $5.9$\%, $8.5$\%, $17.2$\%, and $26.1$\%, showing a monotonic increase with the exception of the elevated value reported for the moderate-$1$ case. It should be noted that while it may be suspected that elevated values of particle residence time ought to appear in the near-wall regions in Fig.\ \ref{fig:EjectionMapa} due to the slower velocities occurring in the boundary layer, the inaptitude of the experimental PIV setup in resolving the flow immediately adjacent to the wall will likely cause the particle residence time to be underestimated in these regions and hence the reported values of $\mathrm{PRT}_2^4$ will also be slightly underestimated in the plane of interest:
			
			\begin{equation}
			\label{eq:PRT24}
				\mathrm{PRT}_2^4 = \frac{1}{A}\iint_A\tau_2^{4*}(x,y)\mathrm{d}A, \quad \tau_2^{4*}(x,y) = \left\lbrace
				\begin{array}{lll}
					0           & & \tau^*(x,y) < \beta + 1 \\
					\tau^*(x,y) & & \beta + 1 \leq \tau^*(x,y) < \beta + 3 \\
					\beta + 3   & & \tau^*(x,y) \geq \beta + 3 \\
				\end{array}\right.
			\end{equation}
			
\begin{figure}[!t]
				\centering
				\subfloat[\label{fig:InjectionMapa}]{%
					\includegraphics[width=0.80\linewidth]{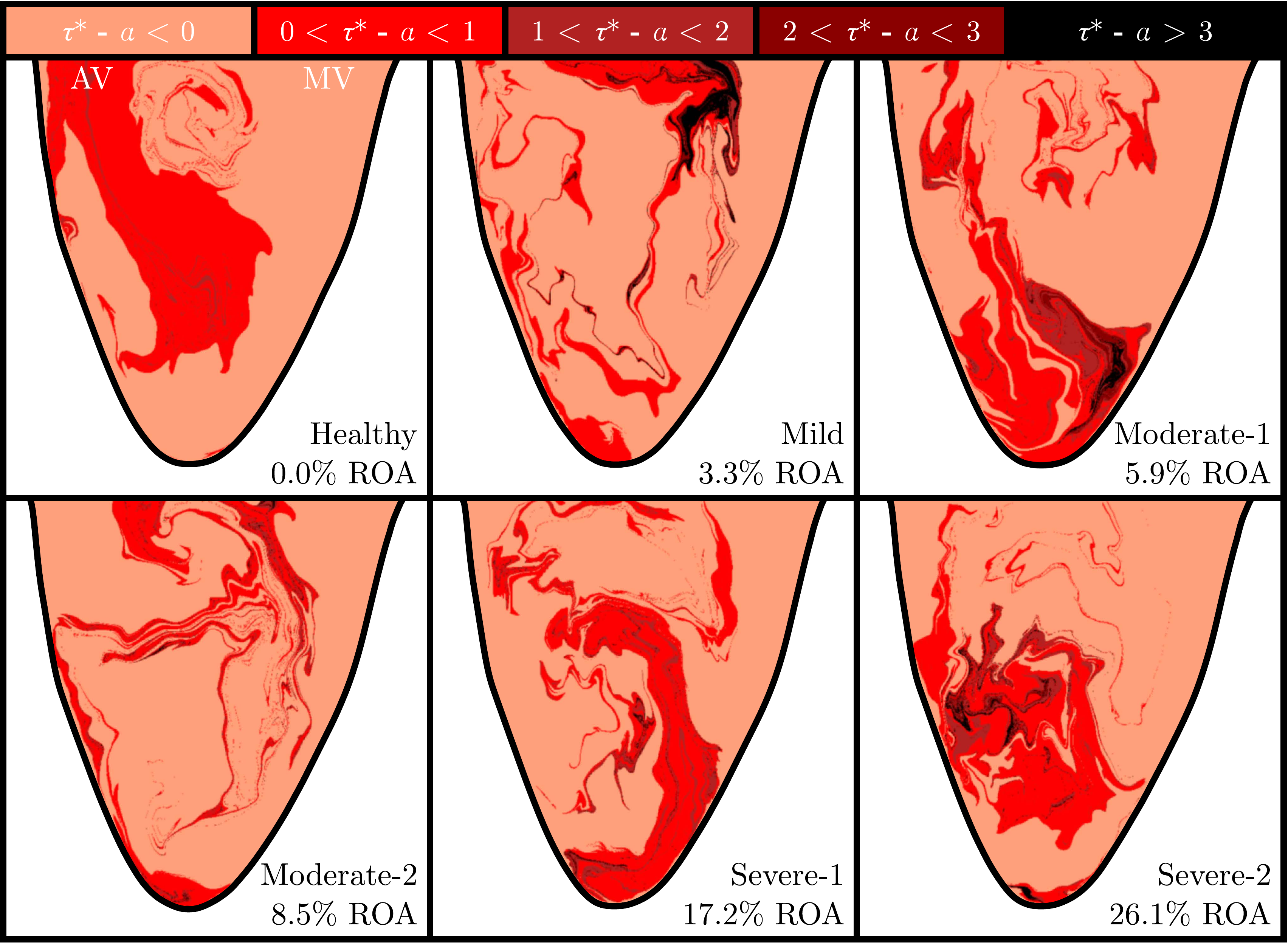}
				}\\
				\subfloat[\label{fig:InjectionMapb}]{%
					\includegraphics[width=0.98\linewidth]{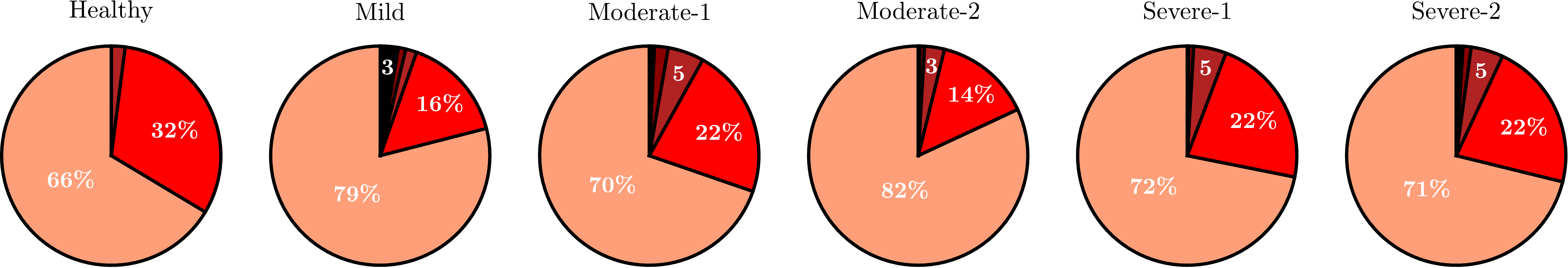}
				}
			\caption{(a) Normalized particle residence time ($\tau^*$) in the left ventricle at the start of the ejection phase ($t^* = 0$) colored by time since injection in beats; $\alpha = 0.562$ is the duration of the filling phase normalized by the cycle period ($T = 0.857$ s). (b) Percentage of ventricle area corresponding to the colored regions in (a).}
			\label{fig:InjectionMap}
			\end{figure}

			In a similar manner to which regions can be delineated by cycle in terms of ejection, if the particles are advected in backward time, the particle residence time could be used to delineate regions of the ventricle by the heart beat during which they have been injected. The results are displayed in Fig.\ \ref{fig:InjectionMap}, computed again at the start of the ejection phase ($t^* = 0$) as for Fig.\ \ref{fig:EjectionMap}. The corresponding percentage of each colored region with respect to the total ventricle area is given in Fig.\ \ref{fig:InjectionMapb}. In the healthy ventricle, a good portion of particles from the previous beat in red are organized for ejection (this region comprises $31.6$\% of the ventricular area). Fluid injected within one beat, indicated by the lightest color in Fig.\ \ref{fig:InjectionMap}, occupies a channel adjacent to the ventricle wall, surrounding fluid injected from the previous beat. Notably, the amount of fluid injected into the left ventricle within one beat is larger in the regurgitant cases than in the healthy case, which is effectively due to the increased inflowing volume characteristic of aortic regurgitation, although the increase is not monotonic here due to three-dimensional effects. A feature that is particularly intriguing in all cases of Fig.\ \ref{fig:InjectionMapa} is the tendency of regions to form thin lamellae, much more so than in the ejection flow map of Fig.\ \ref{fig:EjectionMapa}. Since the left ventricle expands mostly in the out-of-plane direction, the additional inflow from the regurgitation appears as a compressibility in the two-dimensional plane of interest in this study, restricting the remaining fluid area into the observed structures. These lamellae effectively coincide with attracting Lagrangian coherent structures, and provided they persist long enough in the flow, they may manifest as sites at which activated platelets can agglomerate. Furthermore, by referring to both Figs.\ \ref{fig:EjectionMapa} and \ref{fig:InjectionMapa}, a picture of blood stasis can be formed spanning seven cardiac cycles. In this way, it can again be seen that virtually no particles in the healthy scenario reside in the ventricle for four beats (in fact, the largest residence time considering the seven cardiac cycles is only $3.7$ beats). However, by comparing the mild ($\mathrm{ROA}$ equal to $3.3$\%) and moderate-$1$ ($\mathrm{ROA}$ equal to $5.9$\%) cases, there are distinct black regions that clearly overlap, indicating stasis for a total of over seven cardiac cycles. Although it is not clear from comparing Figs.\ \ref{fig:EjectionMapa} and \ref{fig:InjectionMapa}, the moderate-$2$ ($\mathrm{ROA}$ equal to $8.5$\%), severe-$1$ ($\mathrm{ROA}$ equal to $17.2$\%), and severe-$2$ ($\mathrm{ROA}$ equal to $26.1$\%) cases also contain particles residing in the ventricle for over seven cardiac cycles, though considerably less than in the moderate-$1$ case.
			
			\begin{figure}[!t]
				\centering
				\subfloat[\label{fig:DirectFlowMapa}]{%
					\includegraphics[width=0.80\linewidth]{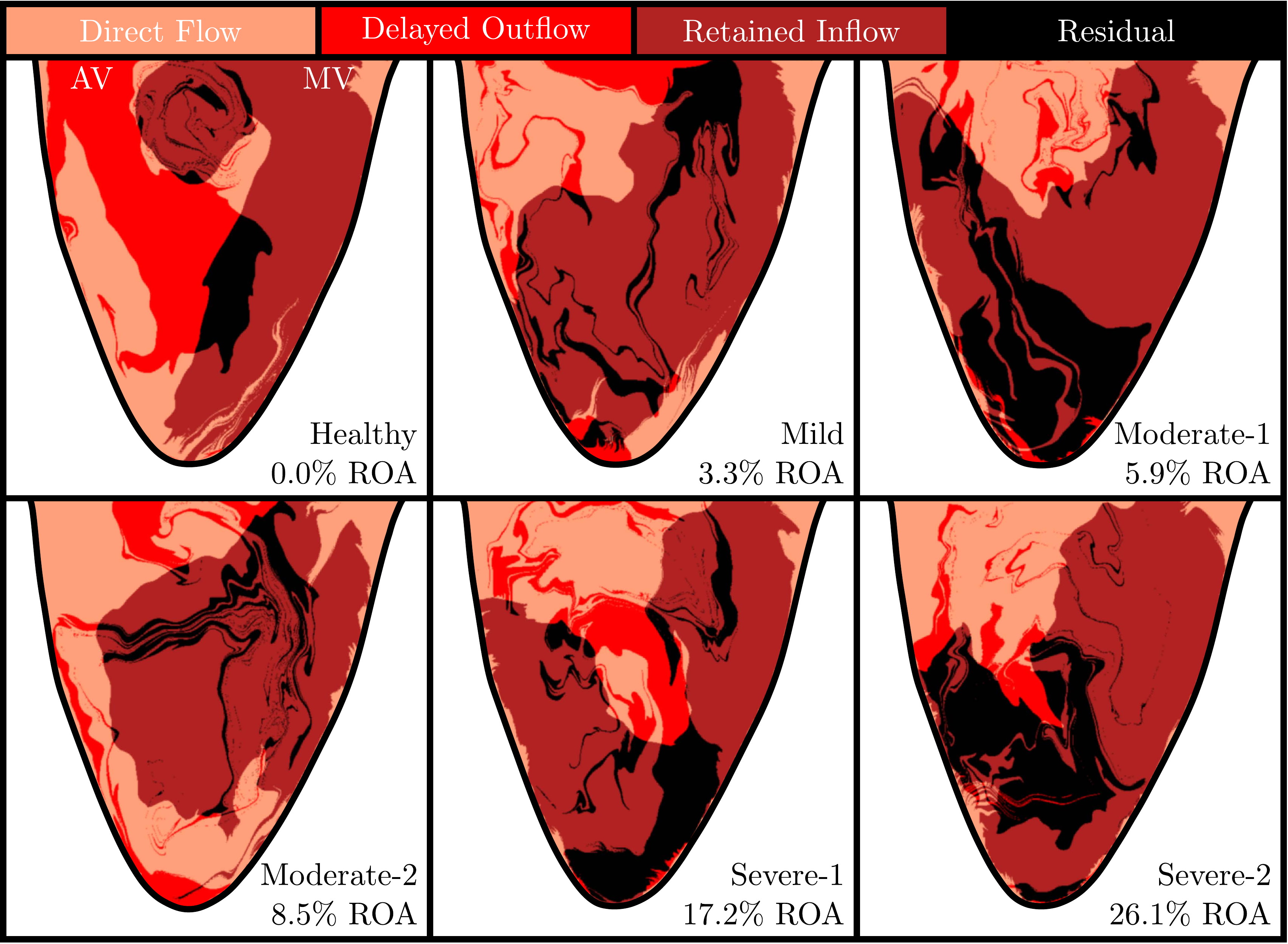}
				}\\
				\subfloat[\label{fig:DirectFlowMapb}]{%
					\includegraphics[width=0.98\linewidth]{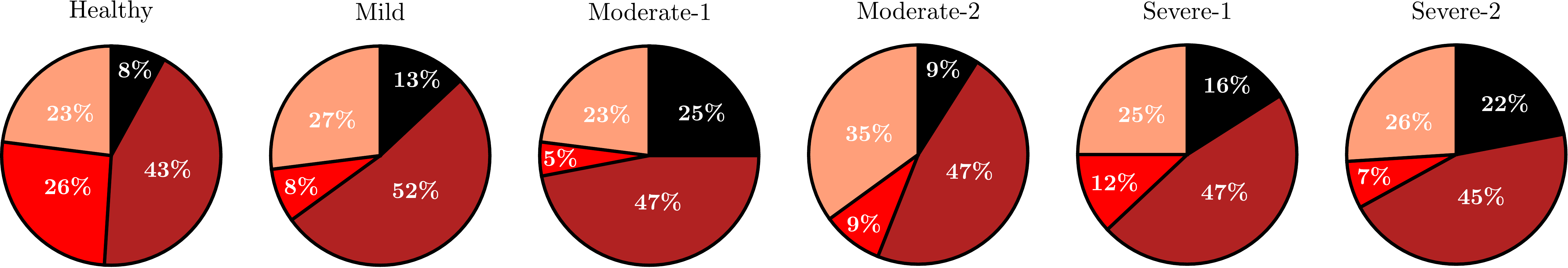}
				}
			\caption{(a) Left ventricular area at the start of the ejection phase ($t^* = 0$) partitioned by blood that ($1$) has entered and ejected the ventricle in the same beat (direct flow), ($2$) was already present in the ventricle and was ejected in the current beat (delayed outflow), ($3$) has entered and remained in the ventricle in the current beat (retained inflow), and ($4$) was already present in the ventricle and not ejected in the current beat (residual volume). (b) Percentage of left ventricular area corresponding to the regions displayed in (a).}
			\label{fig:DirectFlowMap}
			\end{figure}
			
\begin{figure}[!t]
				\centering
				\includegraphics[width=0.80\linewidth]{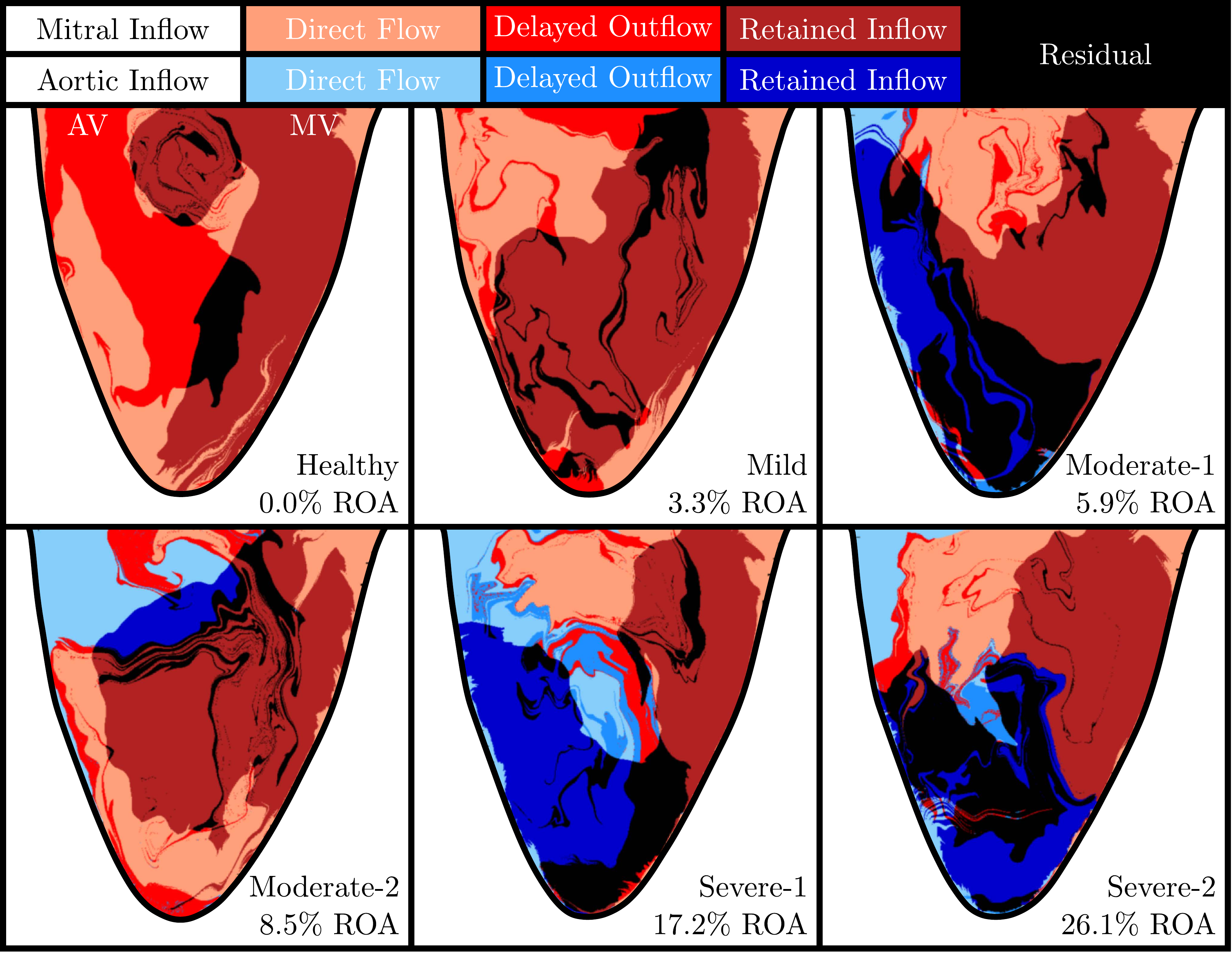}
			\caption{Left ventricular area at the start of the ejection phase ($t^* = 0$) partitioned between aortic (blue) and mitral (red) inflows according to blood that ($1$) has entered and ejected the ventricle in the same beat (direct flow), ($2$) was already present in the ventricle and was ejected in the current beat (delayed outflow), ($3$) has entered and remained in the ventricle in the current beat (retained inflow), and ($4$) was already present in the ventricle and not ejected in the current beat (residual volume).}
			\label{fig:DirectFlowMapReg}
			\end{figure}

			From yet another perspective, the idea of compartmentalizing blood volume into direct flow, delayed ejection flow, retained inflow, and residual volume as in the work of \citeA{Bolger07} is an effective way of describing the pumping efficiency of the left ventricle within a single cycle. This requires Fig.\ \ref{fig:EjectionMap} to be condensed into two simple regions, namely, what will be ejected in the current cycle (beginning from $t^* = 0$) and what will not, and likewise for Fig.\ \ref{fig:InjectionMap}, namely, what has just been injected in the previous cycle (ending at $t^* = 0$) and what has not. Overlapping regions of these plots then partition the ventricular area into the aforementioned compartments. Figure \ref{fig:DirectFlowMapa} displays these partitioned regions for all simulated cases with Fig.\ \ref{fig:DirectFlowMapb} showing the percentage of ventricular area corresponding to each partition. While previous \textit{in vivo} works demonstrated a significant decrease in direct flow in patients with dilated cardiomyopathy \cite{Bolger07, Eriksson13, Hendabadi13}, it appears that the direct flow remains relatively unchanged with aortic regurgitation. This is likely due to any regurgitant inflow remaining in the vicinity of the aortic valve being readily expelled in the following ejection, although, as will be discussed shortly, much of the regurgitant inflow is in fact retained. Characteristically, it seems aortic regurgitation is marked by a significant decrease in delayed ejection flow accompanied by an increase in residual volume. The reduced delayed ejection flow appears to be attributed most to the contraction of material previously present in the ventricle to lamellae-like structures prior to ejection in the plane of interest, dropping from $26$\% in the healthy scenario down to $5$\%-$12$\% for the regurgitant cases. While the results of this flow partitioning show some distinct features, it is clear that for aortic regurgitation it must be reformulated in order to consider two inflows. In this way, there will be a direct flow, delayed ejection flow, and retained inflow for both the regurgitant and mitral jets as well as the residual volume, constituting a total of seven regions. This can be achieved by tracking where the particles flee the top of the flow domain in backward time to determine from which valve they were injected. Here particles were considered to be part of the regurgitant volume if they were ejected (in backward time) through a horizontal segment half the width of the base of the ventricle extending from the wall on the aortic side at the top of the flow domain. This updated flow map is shown in Fig.\ \ref{fig:DirectFlowMapReg}. The most pronounced feature when comparing Figs.\ \ref{fig:DirectFlowMapa} and \ref{fig:DirectFlowMapReg} is that much of the regurgitant inflow is retained (indicated by the dark blue regions). Of the $47.0$\%, $46.8$\%, $46.7$\%, and $44.9$\% of the retained inflow for the moderate-$1$, moderate-$2$, severe-$1$, and severe-$2$ cases displayed in Fig.\ \ref{fig:DirectFlowMapa}, respectively $26.6$\%, $10.9$\%, $50.1$\%, and $28.4$\% are composed of regurgitant inflow. Severe aortic regurgitation (significantly at least in the severe-$1$ case) may be marked by more retained regurgitant inflow. As the regurgitant jet in the case of mild regurgitation could not be seen within the measurement domain, further division of the ventricular area was not possible.
			
			\begin{figure}[!t]
				\centering
				\includegraphics[width=\linewidth]{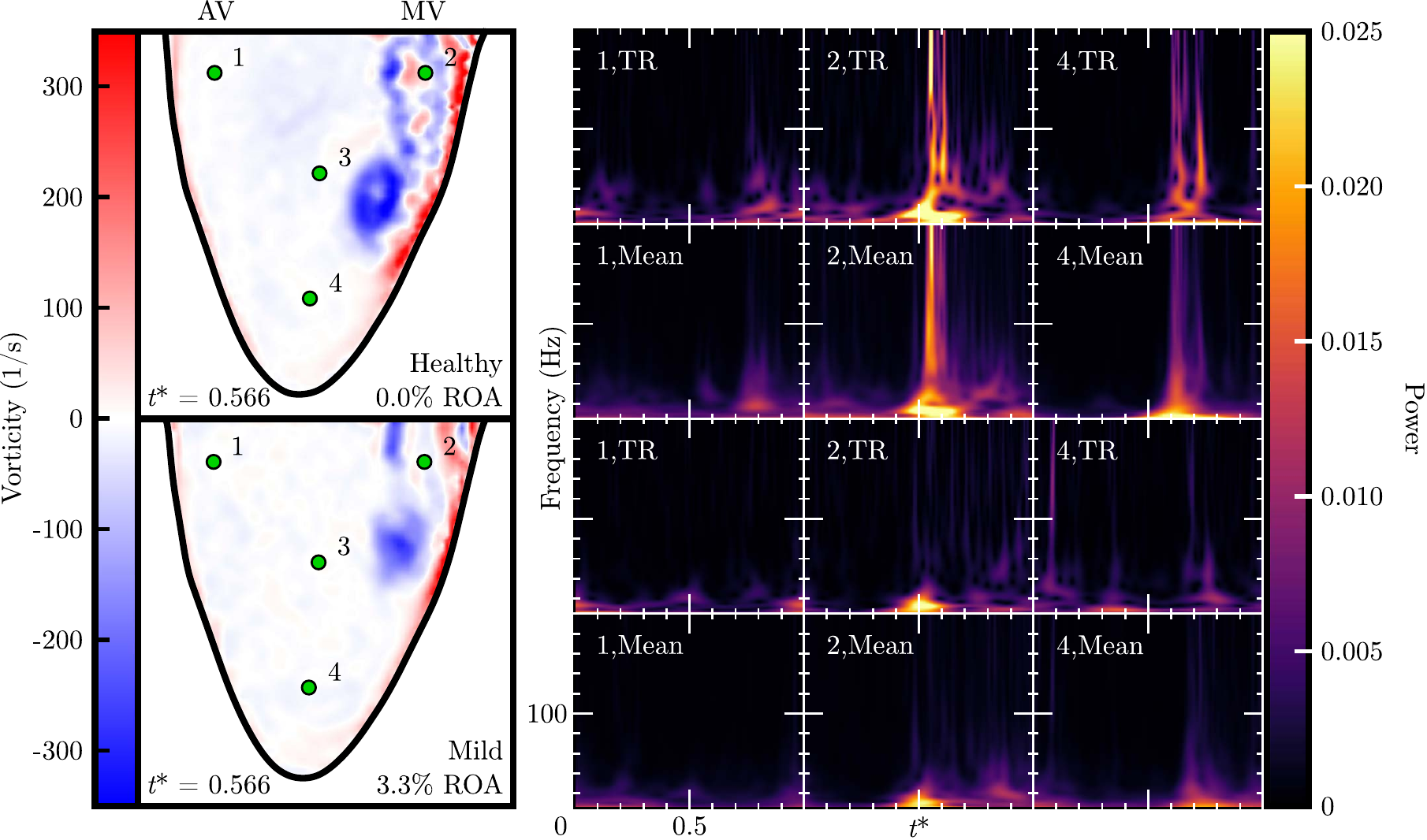}
			\caption{Time-frequency spectra of the $u$ component of velocity (right) at three points for the healthy and mild cases of aortic regurgitation. Here and in the following figures spectra marked with TR denote sample spectra for a single time-resolved velocity signal, while those marked with Mean denote the mean of ten spectra computed from the velocity signal of distinct acquisitions. The top two rows of spectra correspond to the healthy scenario, while the bottom two rows correspond to the mild scenario. The vorticity field (left) is also shown at $t^* = 0.566$, the time at which the highest frequencies are present just downstream of the mitral inflow (point $2$).}
			\label{fig:CWTar01}
			\end{figure}
			
			\begin{figure}[!t]
				\centering
				\includegraphics[width=\linewidth]{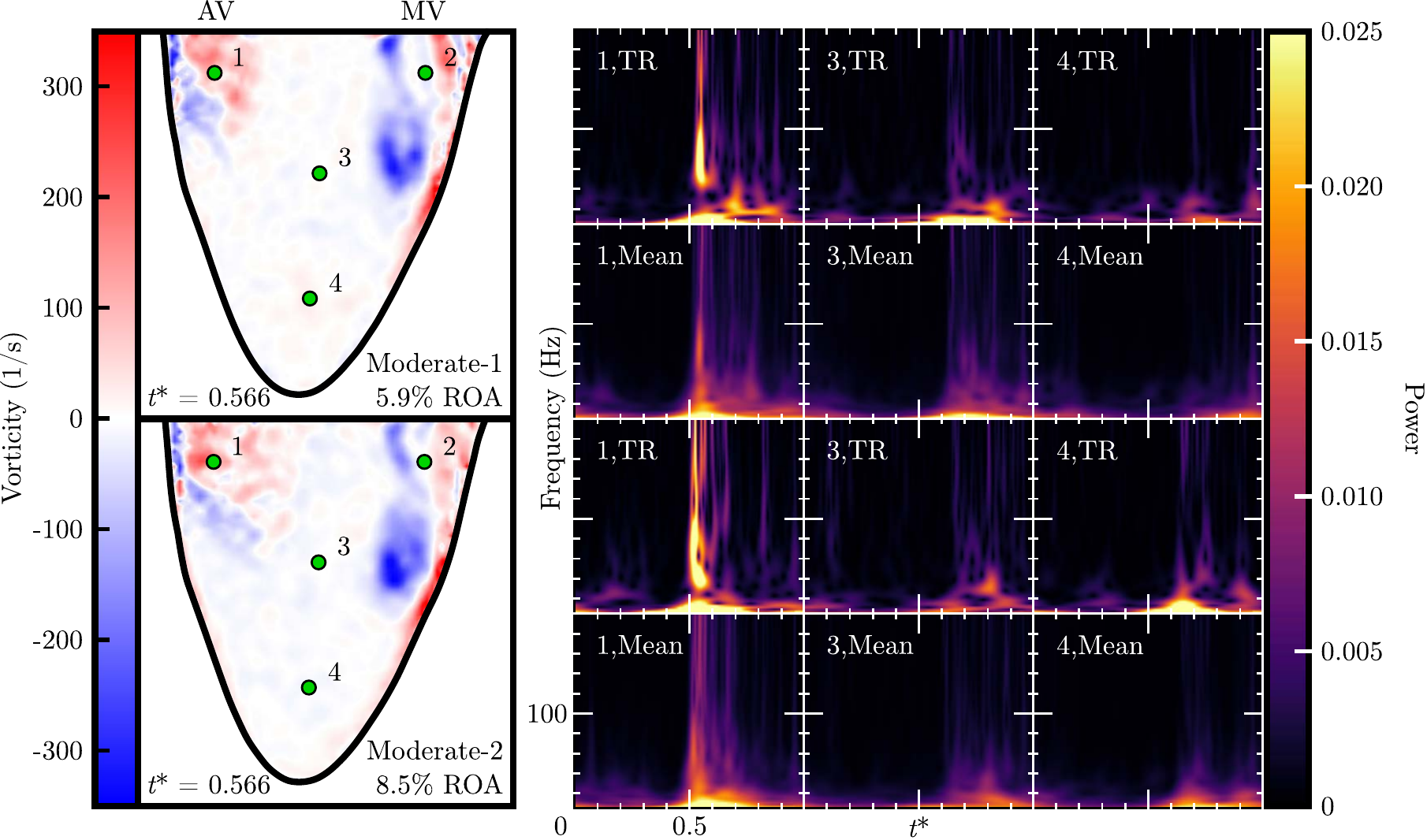}
			\caption{Time-frequency spectra of the $u$ component of velocity (right) at three points for the moderate cases of aortic regurgitation. The top two rows of spectra correspond to the moderate-$1$ scenario, while the bottom two rows correspond to the moderate-$2$ scenario. The vorticity field (left) is also shown at $t^* = 0.566$.}
			\label{fig:CWTar23}
			\end{figure}
			
\begin{figure}[!t]
				\centering
				\includegraphics[width=\linewidth]{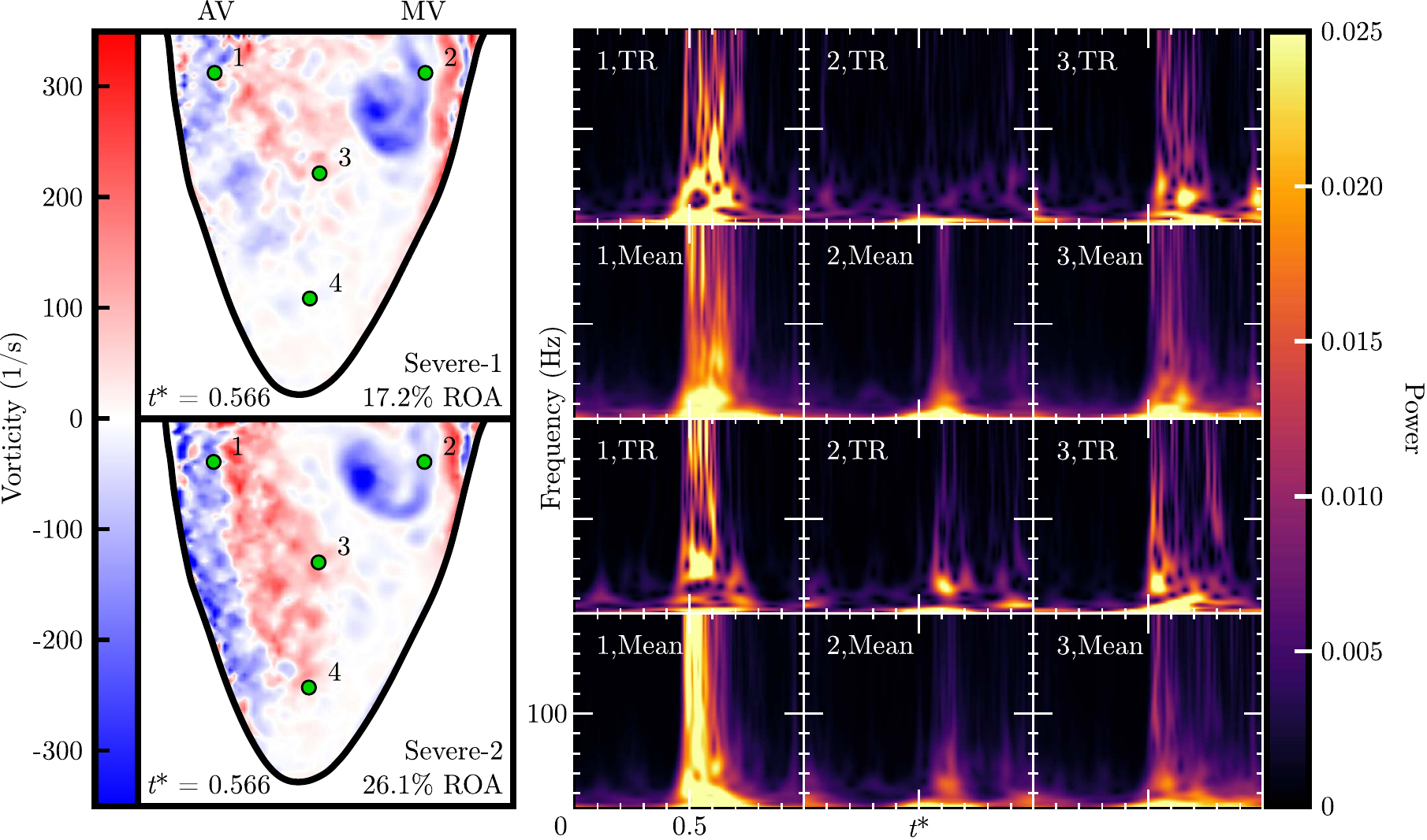}
			\caption{Time-frequency spectra of the $u$ component of velocity (right) at three points for the severe cases of aortic regurgitation. The top two rows of spectra correspond to the severe-$1$ scenario, while the bottom two rows correspond to the severe-$2$ scenario. The vorticity field (left) is also shown at $t^* = 0.566$.}
			\label{fig:CWTar45}
			\end{figure}

			The occurrence of increased stasis in the moderate-$1$ case seen in Figs.\ \ref{fig:EjectionMap}–\ref{fig:DirectFlowMapReg} as compared to the three more severe cases may be a direct result of laminar versus intermittently turbulent activity within the left ventricle. In order to investigate this idea further, the velocity signals at four points within the left ventricle were analyzed using their time-frequency spectra computed with the continuous wavelet transform (using the complex Morlet wavelet) for all cases. One point was placed just downstream of each jet (mitral and regurgitant) to observe their degree of intermittency. Additionally, judging by the particle residence time mapped in Figs.\ \ref{fig:EjectionMap} and \ref{fig:InjectionMap}, one point was placed in the ventricle center and another in the apex. It should be noted that the spectra were computed for both the $u$ and $v$ components of the velocity signals at each of the four points of interest for all cases, however the resulting time-frequency spectra effectively provide very similar qualitative and quantitative insight, therefore we only discuss the $u$ component of velocity here. The spectra were computed for all ten recorded time-resolved signals at each point. In Figs.\ \ref{fig:CWTar01}-\ref{fig:CWTar45}, those spectra marked with TR denote sample spectra for a single time-resolved velocity signal, while those marked with Mean denote the mean of the ten corresponding spectra, giving some indication of the repeatability of the observed frequencies in the experiments. Additionally, the three figures were all constructed using the same scales for the sake of ease of comparison. The respective vorticity fields at $t^* = 0.566$, the time at which the highest frequencies are present just downstream of the mitral inflow (point $2$), are also shown. The velocity gradients used to obtain the vorticity were computed using the fourth-order, noise-optimized hybrid compact-Richardson scheme presented in \citeA{Etebari05}, which was shown to have lower noise amplification than the second-order explicit finite-difference scheme while also reducing bias error in the calculation of vorticity from PIV data. Figure \ref{fig:CWTar01} shows the time-frequency spectra for the velocity signals at three of the four points for both the healthy and mild ($\mathrm{ROA}$ equal to $3.3$\%) cases. It is clear that in the vicinity of the aortic valve (point $1$), there is little to no high frequency present in the velocity signals, mainly because it is not significantly disturbed by regurgitation in both cases. The same is true in the ventricle center which, although not shown here, resembles the spectra of point $1$. Downstream of the mitral valve (point $2$), high frequencies are observed for a very brief period during the \textit{E} wave of filling for the healthy case, while this effect seems to be largely suppressed with mild aortic regurgitation. This brief intermittency in the healthy case can be seen rather clearly in its respective vorticity field in Fig.\ \ref{fig:CWTar01}, whereas it is very faint in the vorticity field for mild regurgitation. The same high frequencies are seen in the ventricle apex (point $4$) at a slightly later time, namely, once the mitral inflow penetrates deep enough downstream. It appears that the entrainment of the regurgitant volume by the mitral vortex with mild regurgitation diminishes the inflow jet velocity just enough in early filling to virtually eliminate the intermittency; however, as previously discussed, the increased stasis is largely due to the resulting dynamics of the mitral vortex rather than intermittent activity. The two moderate cases of regurgitation ($\mathrm{ROA}$ equal to $5.9$\% and $8.5$\%) are compared in Fig.\ \ref{fig:CWTar23}, this time the focus being on the point in the vicinity of the aortic valve (point $1$) and the two points within the ventricle (points $3$ and $4$). The regurgitant jet in all cases more severe than mild regurgitation has associated with it prolonged intermittency during left ventricular filling, evidenced by the vivid occurrence of high frequencies in the time-resolved and mean spectra for point $1$ and the corresponding vorticity fields in Figs.\ \ref{fig:CWTar23} and \ref{fig:CWTar45}. On the contrary, the mitral inflow appears to remain distinctly laminar with regurgitation, showing very weak high-frequency contributions in the spectra. Although not shown in Fig.\ \ref{fig:CWTar23}, the spectra at the point just downstream of the mitral valve (point $2$) for both moderate cases exhibit behavior similar to that of the mild case in Fig.\ \ref{fig:CWTar01}. Most interesting between the two moderate cases of regurgitation is the behavior at points $3$ and $4$. While a similar degree of intermittency can be observed in the ventricle center (point $3$), fluid motion in the ventricle apex (point $4$) is significantly more intermittent in the moderate-$2$ case than for the moderate-$1$ case, evidenced by both the magnitude of the observed high frequencies and the duration over which they exist. The mean of the spectra in the apex in the moderate-$2$ case demonstrates the persistent occurrence of these high frequencies over multiple cycles. Mixing in the apex is therefore further deterred in the moderate-$1$ case in part from a lack of intermittency, giving rise to locally increased stasis. In the two severe cases ($\mathrm{ROA}$ equal to $17.2$\% and $26.1$\%), the flow is largely intermittent within the entire ventricular volume, a feature qualitatively clear from the vorticity fields in Fig.\ \ref{fig:CWTar45}. The regurgitant jet is clearly turbulent throughout its propagation and exhibits significant high-frequency contributions in the spectra at point $1$ just downstream of the aortic valve. The turbulent propagation of the jet is also seen in the ventricle center (point $3$) and apex (point $4$), where the spectra in the apex, although not shown, are similar to those in the center with lesser magnitude. Additionally, the brief intermittency observed in the spectra just downstream of the mitral valve (point $2$) returns in the severe cases, although in this case the turbulent regurgitant jet likely influences these high frequencies as well.
			
		\subsection{\label{sec:LCSs}Lagrangian coherent structures and aortic regurgitation}
		
			The finite-time Lyapunov exponent is a measure of the maximum exponential rate of separation of initially very close particles over a finite time interval. Briefly, given some particle initially located at position $\mathbf{x}_0$ at time $t_0$, after some time $T$ it will be advected to a new location $\mathbf{x}$ defined by some flow map, say, $\mathbf{F}_{t_0}^{t_0 + T}$, i.e.,
			\begin{equation}
			\label{eq:FlowMap}
				\mathbf{x}_0 \mapsto \mathbf{F}_{t_0}^{t_0 + T}\left(\mathbf{x}_0\right) = \mathbf{x}\left(t_0 + T;t_0,\mathbf{x}_0\right).
			\end{equation}
			Now, as described in \citeA{Shadden05}, for two initially infinitesimally close particles separated by a distance $\delta \mathbf{x}\left(t_0\right)$, maximal stretching between the two particles after advection from the same flow map $\mathbf{F}_{t_0}^{t_0 + T}$ will occur when the initial particle separation is aligned with the eigenvector associated with the maximum eigenvalue of the right Cauchy-Green deformation tensor, i.e.,
			\begin{equation}
			\label{eq:MaxSep}
				\max_{\delta\mathbf{x}\left(t_0\right)}||\delta\mathbf{x}\left(t_0 + T\right)|| = \sqrt{\lambda_{\mathrm{max}}\left(\mathbf{C}_{t_0}^{t_0 + T}\left(\mathbf{x}\right)\right)}||\overline{\delta \mathbf{x}\left(t_0\right)}|| = e^{\sigma_{t_0}^{|T|}\left(\mathbf{x}\right)|T|}||\overline{\delta \mathbf{x}\left(t_0\right)}||,
			\end{equation}
			with $\overline{\delta \mathbf{x}\left(t_0\right)}$ denoting the eigenvector and where the right Cauchy-Green deformation tensor is given by
			\begin{equation}
			\label{eq:CauchyGreen}
				\mathbf{C}_{t_0}^{t_0 + T}\left(\mathbf{x}\right) = \left[\nabla\mathbf{F}_{t_0}^{t_0 + T}\left(\mathbf{x}\right)\right]^*\nabla\mathbf{F}_{t_0}^{t_0 + T}\left(\mathbf{x}\right)
			\end{equation}
			and
			\begin{equation}
			\label{eq:FTLE}
				\sigma_{t_0}^{|T|}\left(\mathbf{x}\right) = \frac{1}{2|T|}\ln\left(\lambda_{\mathrm{max}}\left(\mathbf{C}_{t_0}^{t_0 + T}\left(\mathbf{x}\right)\right)\right)
			\end{equation}
			is defined as the finite-time Lyapunov exponent. Finite-time Lyapunov exponents can be computed in forward or backward time, where measuring particle separation in backward time is equivalent to particle convergence in forward time. Forward and backward FTLE fields are here employed for their superior flow visualization properties and to heuristically locate, track, and analyze the respective repelling and attracting LCSs from the FTLE ridges arising under the studied experimental flow conditions.
			
			The FTLE fields were computed in both forward and backward time using the classical fourth-order Runge-Kutta scheme for particle advection (using bicubic interpolation to acquire velocities at the advected particle positions) and an explicit second-order centered finite-difference scheme to compute the derivatives required to construct the right Cauchy-Green deformation tensor. The integration was carried out over two cardiac cycles forward and backward, similar to \citeA{Hendabadi13}. Evidently, the longer the selected period of advection, the more virtual particles ought to exit the flow domain. Here the FTLEs at the initial positions of particles that have fled the flow domain were held as their final values before fleeing. Notably, there are other ways such a problem can be treated; for instance, \citeA{Hendabadi13} continued to advect the fleeing particles at a constant velocity equal to their respective exiting velocities, although without knowledge of the velocity field outside the flow domain, there is no consensus on how to treat this condition. In order to obtain sharper LCSs in the FTLE fields, the virtual particles were seeded in an eightfold refined Cartesian grid. As illustrated in \citeA{Vetel09}, refining the grid has the effect of sharpening the detail in the FTLE field, particularly its ridges, and permits a longer integration time as more particles ought to remain in the flow domain for longer periods. There was little difference in the detail revealed by $16$-fold and eightfold refined grids for the selected two-cycle integration time, therefore the latter refinement was chosen.
			
			\begin{figure}[!t]
				\centering
				\includegraphics[width=1.00\linewidth]{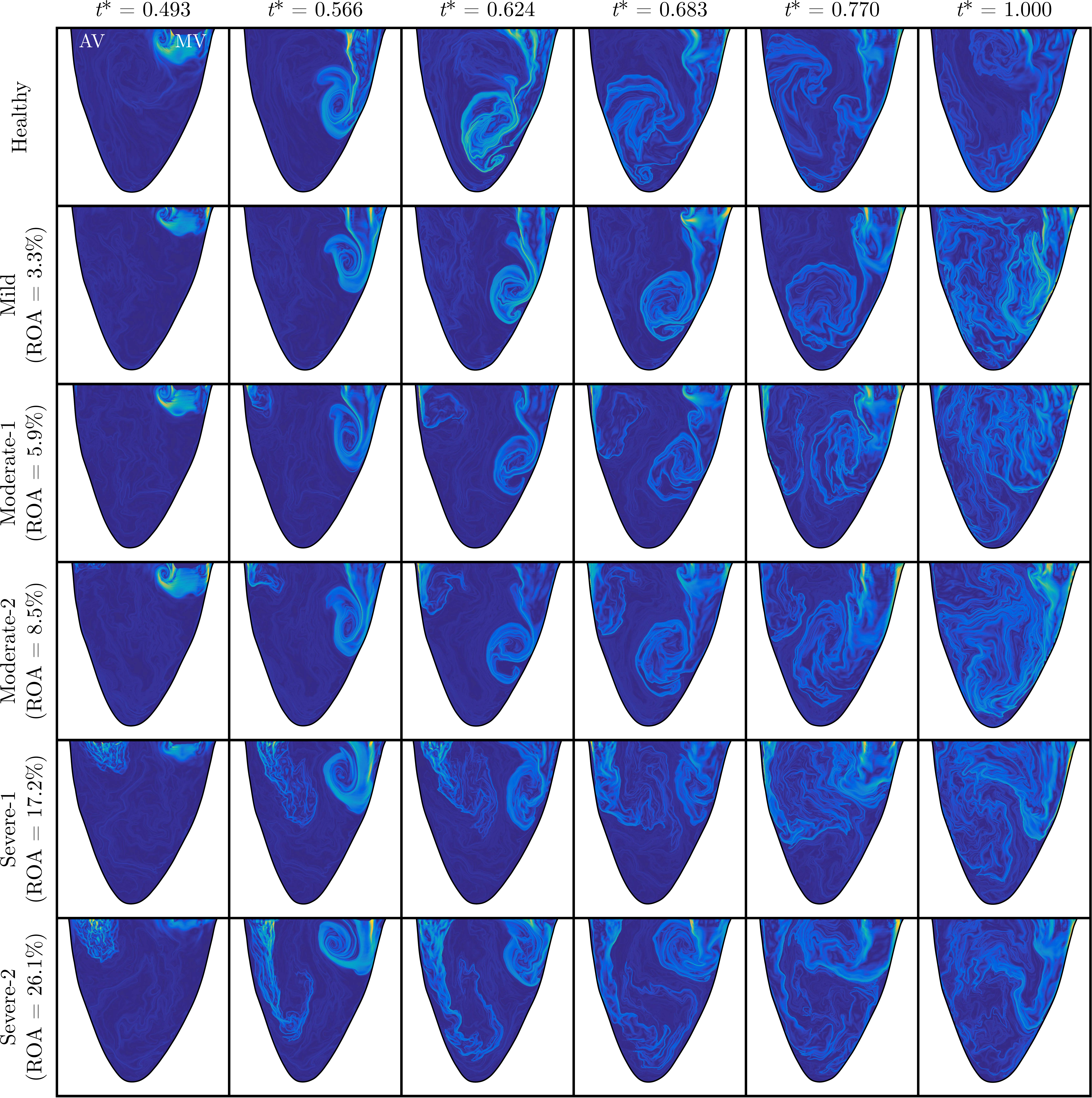}
			\caption{Backward FTLE fields for six severities of aortic regurgitation (including the healthy scenario) at the same six time instants throughout the filling phase of the left ventricle. The backward FTLE fields were computed over an integration time of two cycles.}
			\label{fig:FTLEbwd}
			\end{figure}
			
			\begin{figure}[!t]
				\centering
				\includegraphics[width=1.00\linewidth]{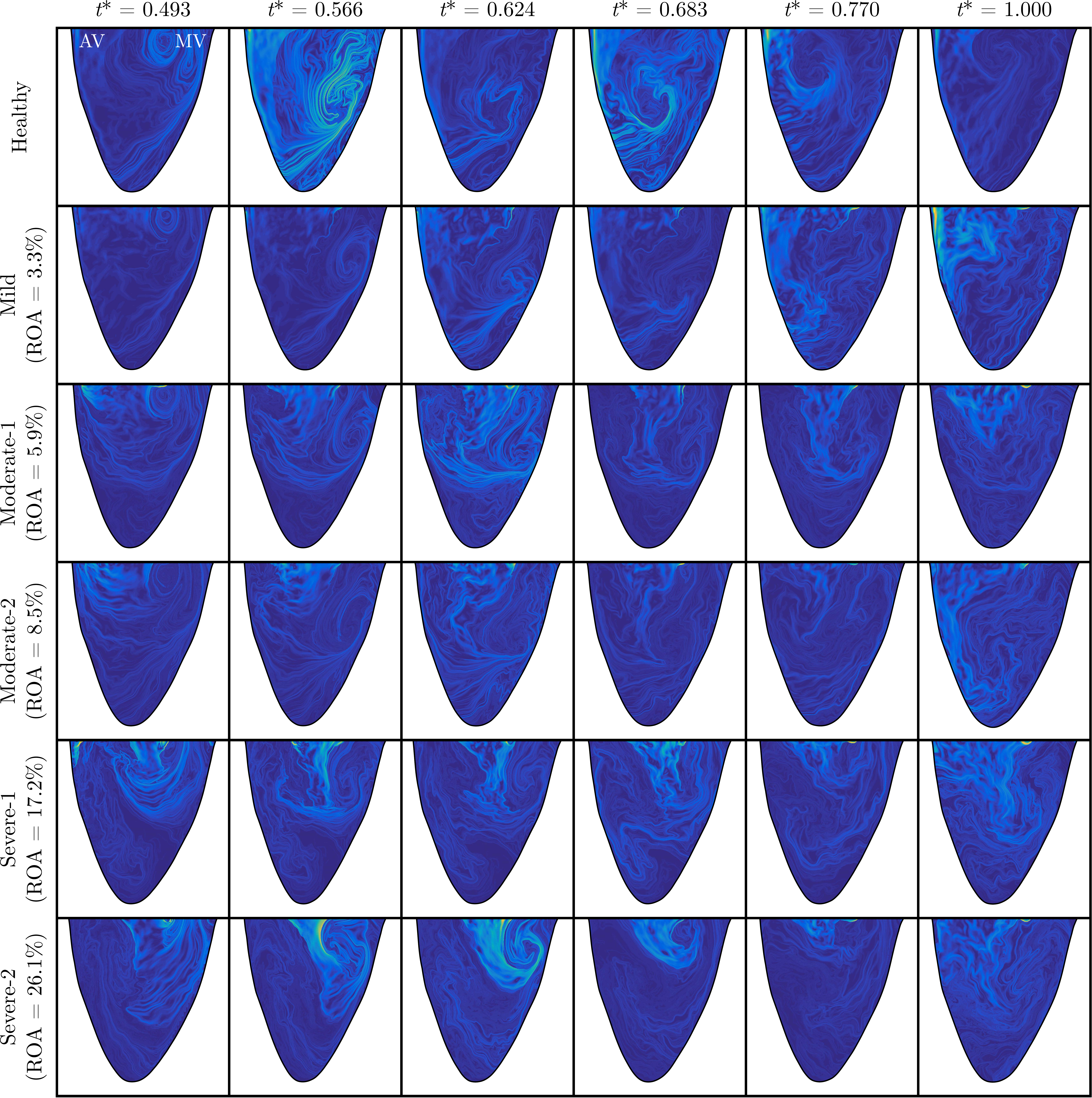}
			\caption{Forward FTLE fields for six severities of aortic regurgitation (including the healthy scenario) at the same six time instants throughout the filling phase of the left ventricle. The forward FTLE fields were computed over an integration time of two cycles.}
			\label{fig:FTLEfwd}
			\end{figure}
			
			Figure \ref{fig:FTLEbwd} shows the evolution of the backward FTLE field for all simulated cases (the corresponding videos over a complete cycle are available in the Supplemental Material). In the case of the healthy left ventricle (Fig.\ \ref{fig:FTLEbwd}, first row), the backward FTLE field shows a concentrated vortex forming at the start of filling as an attracting material structure, entraining mostly inflowing fluid. The generated vortex forms a channel with the ventricle wall on the mitral side, directing inflowing blood from the mitral valve down to the ventricle apex. The channel front is itself a ridge of the FTLE field as it displaces fluid to make way for the inflow, effectively forming a transport barrier between the two fluid volumes. Hence, as discussed in \citeA{Hendabadi13}, there is an attracting LCS bounding the volume of the mitral inflow. In the same manner, there is a repelling LCS bounding the volume of fluid that will be ejected in the forward FTLE fields. This vortex core propagates further downstream into the left ventricle, extending the channel until the vortex core approaches close enough to the ventricle wall, where it is forced to follow a curved trajectory away from the wall, finally allowing the mitral inflow jet to expand. This delayed expansion of the mitral inflow has also been observed and described \textit{in vivo} in \citeA{Charonko13}. From the backward FTLE field, it becomes evident that transport of inflowing blood largely occurs adjacent to the contour of the ventricle wall for a healthy left ventricle, having a washing effect for blood in the vicinity of the ventricle apex; this can also be seen in Fig.\ \ref{fig:InjectionMapa}. By the end of the filling phase ($t^* = 1$), the clockwise vortex has set up in the ventricle's center and much of the injected blood occupies a thick margin alongside the ventricle wall. Together this vortex and inflow organize and bound blood previously present in the left ventricle for ejection. With regard to aortic regurgitation, we start by noting some key features described previously that are immediately evident from the backward FTLE field. For instance, in the first column of Fig.\ \ref{fig:FTLEbwd} ($t^* = 0.493$), representing just a few moments after the start of filling, there is some indication about the timing between the onsets of the mitral and regurgitant jets. The regurgitant jet has not yet emerged in the moderate-$1$ case ($\mathrm{ROA}$ equal to $5.9$\%) while it has just begun to emerge in the moderate-$2$ case ($\mathrm{ROA}$ equal to $8.5$\%) and has clearly emerged prior to the mitral jet in the two severe cases ($\mathrm{ROA}$ equal to $17.2$\% and $26.1$\%). It appears that the timing between the onset of the mitral and regurgitant jets may lead to significant differences in overall particle residence times while still resulting in the monotonic increase in the total energy being dissipated by viscous forces per cycle with regurgitation severity, as shown in our previous work \cite{DiLabbioKadem18}. While the regurgitant jets in all cases appear to drive the mitral vortex toward the ventricle wall (seen in the second column of Fig.\ \ref{fig:FTLEbwd}), the timing at which it occurs in the moderate-$1$ case ($\mathrm{ROA}$ equal to $5.9$\%) appears to be most detrimental in the sense of blood stasis. The difference between the moderate-$1$ and moderate-$2$ cases is made clear in the fifth column of Fig.\ \ref{fig:FTLEbwd} ($t^* = 0.770$), where in the moderate-$1$ case the regurgitant jet managed to match the penetration depth of the mitral jet whereas in the moderate-$2$ case the regurgitant jet has not penetrated as deep and opposes the ascent of the mitral vortex. The difference at the end of the filling phase between the two is that in the moderate-$1$ case a weak counterclockwise vortex sets up in ventricle apex and in the moderate-$2$ case a weak counterclockwise vortex sets up just below the aortic valve. With severe regurgitation, the regurgitant jet begins early and quickly overtakes the mitral jet, restricting downward progression of the mitral vortex and thus forcing it to remain in close proximity to the mitral valve. The final penetration depth of the mitral inflow can be seen by a bounding ridge in the last two columns of Fig.\ \ref{fig:FTLEbwd}, an attractive barrier where the regurgitant and mitral inflow volumes clash. In addition to timing, this \textit{in vitro} model suggests that the regurgitant jet is marked by distinct dynamics associated with the severity of the regurgitation. For instance, with moderate regurgitation, the jet begins slowly and gradually propagates to some extent downstream into the ventricle. In contrast, with severe regurgitation, the jet very rapidly propagates into the ventricle during the \textit{E} wave of filling and then quickly becomes less forceful as the pressure gradient across the aortic valve equilibrates.
			
			Rather interestingly, from the second column of Fig.\ \ref{fig:FTLEbwd} ($t^* = 0.566$), it can be seen that the regurgitant jet is not accompanied by a clear vortical structure in this Lagrangian picture whereas the mitral inflow consistently reproduces a clear and well-behaved vortex during early filling, prior to expansion or interaction with the regurgitant inflow. The absence of a visible vortex pattern accompanying the regurgitant jet suggests that the corresponding vortex observed in the Eulerian picture is a poorly attracting structure (as previously evidenced by its intermittency), an indication of its inability to entrain and transport surrounding material. Such a property may in fact be rather beneficial, having a washing effect for blood in the vicinity of the ventricle apex (as is observed in the moderate-$2$ and severe cases) rather than trapping blood in a coherent counterclockwise laminar spiral (as occurs in the moderate-$1$ case). Additionally, the mitral inflow seems to be consistently marked by smooth spiraling ridges, while the regurgitant inflow possesses a somewhat chaotic and broken group of attracting material lines. Nonetheless, in the last column of Fig.\ \ref{fig:FTLEbwd}, marking the end of the filling phase, the presence of a large number of convoluted attracting structures is clear for all cases when compared to the healthy left ventricle. These structures persist throughout the next cycle and can be faintly observed in columns $1$-$5$, effectively presenting sustained locations in the left ventricle with favorable conditions for activated platelets to agglomerate.
			
			The finite-time Lyapunov exponent computed in forward time, displayed in Fig.\ \ref{fig:FTLEfwd}, reveals the repelling material structures in the flow. Unlike attracting material structures from the backward FTLE fields, which manifest as directly tangible structures in the flow, the repelling structures represent hidden organizing features. Additionally, repelling material structures have been suggested as activation sites for platelets in \citeA{Shadden13}, where the authors defined a platelet activation potential and demonstrated their coincidence with repelling material ridges of the FTLE field. In general, one may suspect therefore that platelets are activated in the vicinity of repelling material lines and coalesce along attracting material lines, however no inference is made here regarding activation of the platelets themselves in the case of aortic regurgitation. Perhaps the most notable feature in the forward FTLE fields presented in Fig.\ \ref{fig:FTLEfwd} is the appearance of a near-horizontal repelling ridge in the ventricle in the moderate-$1$, moderate-$2$, and severe-$1$ cases, separating the apical region from the base of the ventricle. While this horizontal barrier is weak in magnitude, it suggests some quiescence of blood in the apical region during the early filling phase. Interestingly, in the severe-$2$ case, at $t^* = 0.493$, it can be seen that a repelling ridge engulfs the mitral region, limiting its propagation downstream early on. This repelling structure shrinks, further restricting the mitral inflow until ultimately releasing a pathway adjacent to the ventricle wall on the mitral side (as seen at $t^* = 0.683$), allowing the mitral inflow to propagate further downstream.
			
	\section{\label{sec:Conc}Conclusion}
		
		With valvular heart diseases posing an escalating burden on the world's aging population, there is a need to study these diseases further in the hopes of detecting and treating them early on. In particular, aortic regurgitation presents an interesting fluid dynamics problem, akin to a confined elastic vessel filling from two impulsive jets. In the present study, aortic regurgitation was simulated \textit{in vitro} on a double activation left heart duplicator, taking into account the physiological adaptations observed \textit{in vivo} in chronic (or gradually occurring) regurgitation. While our previous work focused on a fundamental Eulerian flow description as well as viscous energy loss characteristics, the present work investigated the material transport characteristics from a purely Lagrangian perspective.
		
		The \textit{in vitro} model used in this work further supports the body of literature regarding the optimal nature of a healthy left ventricular flow. The tendency of the mitral vortex to set up in the ventricle's center to impart a single coherent clockwise motion to the entire ventricular fluid volume organizes material for ejection in a near-optimal manner. In a healthy left ventricle, material at the beginning of the filling phase was shown to organize into a columnar region under the left ventricle outflow tract just prior to ejection, permitting a healthy ventricle to be capable of ejecting half its fluid volume in a single beat with minimal viscous energy loss. Within two beats, nearly all material in the left ventricle is cleared ($96$\% in this study), highlighting its performance in maintaining a low overall particle residence time. Inflowing blood was found to be transported mainly along a channel adjacent to the ventricle wall, surrounding blood previously present in the left ventricle and having an important washing effect for particles in the apical region. The backward finite-time Lyapunov field reveals how the vortex generated from the mitral inflow represents an attracting material structure, entraining and transporting fluid downstream into the left ventricle, forming a channel with the ventricle wall, and delaying the expansion of the mitral jet.
		
		With aortic regurgitation, the double-jet filling pattern significantly alters material transport behavior within the left ventricle. It was shown that even mild regurgitation is enough to perturb material transport dramatically, inadequately organizing material into a columnar region under the left ventricle outflow tract. Through an investigation of the particle residence time, it was observed that moderate regurgitation permits the largest degree of blood stasis. This result was largely influenced by the timing between the onset of the regurgitant and mitral jets and decreased intermittently turbulent activity in the apical region. While it may seem that an additional washing effect in the ventricle apex arises with higher regurgitation severities, the viscous energy loss still does increase, as shown in our previous work, implying a larger work input requirement on the part of the heart muscle to eject blood. Interestingly, however, all simulated cases of aortic regurgitation contained particles residing in the left ventricle for over seven cardiac cycles, while no particles remain for over four cycles in the healthy left ventricle. By partitioning the left ventricular volume into direct flow, delayed ejection flow, retained inflow, and residual volume, it was observed that aortic regurgitation could be characterized by a significant drop in delayed ejection flow accompanied by an increase in residual volume. While this provides a rather interesting Lagrangian picture of the flow, it does not distinguish regions of particles by whether they have been injected from the aortic or mitral valve. Such a map was also provided in this work, demonstrating that while any regurgitant volume remaining in the vicinity of the aortic valve is readily ejected, much of the regurgitant volume is in fact retained. The subsequent interaction and clashing of the regurgitant and mitral volumes was more clearly depicted in the backward finite-time Lyapunov exponent fields. Where the two volumes meet, particularly in the severe cases, a sharp transport barrier exists, limiting the progression of the mitral inflow volume in the plane. By the end of the filling phase, the cases exhibiting aortic regurgitation are marked by many remnant attracting ridges of the backward FTLE field. This result outlines a particularly important outcome of this study, namely, that there is a clear increase in blood stasis in the left ventricle associated with aortic regurgitation in addition to sustained locations where activated platelets can coalesce. In other words, aortic regurgitation presents ideal conditions in the left ventricle for the formation of mobile platelet aggregates, however small, which can form clots elsewhere in the body when ejected from the ventricle. In combination with comorbidities that are known to activate platelets, chronic aortic regurgitation poses an additional risk factor, particularly in its moderate stage.
		
		It is to be noted that, without a doubt, this study comes with some important limitations, the most prominent of which is the description of an inherently three-dimensional flow using a single two-dimensional plane. The two-dimensional plane used in this study was specifically selected based on the plane typically used in clinical practice to assess aortic regurgitation via ultrasound. Throughout this work, it was made clear that a three-dimensional description is necessary in view of several results, particularly due to the contraction of areas seen in the particle residence time maps (specifically the map distinguishing areas by the cycle in which they were injected). In fact, proper particle advection in general requires the full three-dimensional flow since it is highly unlikely that particles will remain in the measurement plane for the entire advection period. Additionally, while we have used a trileaflet valve in the mitral position for this work, a bileaflet valve better represents left ventricular physiology. Another limitation, as we have alluded to in Sec.\ \hyperref[sec:BloodTrans]{\ref*{sec:ResDisc}.\ref*{sec:BloodTrans}}, is that the lack of resolution in the near-wall region of the left ventricle likely underestimates particle residence times. This lack of resolution is intrinsic to the methodology used and is better acquired using direct numerical simulation, for which the results of this work may serve as validation away from the wall. The reported values of $\mathrm{PRT}_2^4$ may therefore be considered as a lower bound in the plane of interest, though they are not suspected to change significantly as the area in the near-wall region is small compared to the rest of the ventricular area. The particle residence time maps shown in Figs.\ \ref{fig:EjectionMapa} and \ref{fig:InjectionMapa} can consequently display longer stasis in the vicinity of the wall and this only further highlights the possibility of thrombus formation in the case of aortic regurgitation. Yet another important limitation which we have only briefly alluded to in Sec.\ \hyperref[sec:BloodTrans]{\ref*{sec:ResDisc}.\ref*{sec:BloodTrans}} is the contractile dynamics of the experimental model left ventricle with reference to the physiological contraction of the heart muscle. It is well known that the orientation of the muscle fibers surrounding the left ventricle result in a twisting action of its walls during its ejection phase followed by a rapid untwisting action as the heart muscle relaxes for the filling phase. In the heart simulator used throughout this work, as is currently the case for all such PIV-compatible simulators, this twisting action is entirely neglected and the left ventricle consequently contracts primarily in the radial direction. Evidently, the swirling motion of the flow in the left ventricle is three dimensional and this twist likely promotes this behavior. Inclusion of such structural dynamics is therefore critical in studying the three-dimensional flow and is especially important in resolving the near-wall behavior of the flow. As this work deals with a single two-dimensional plane, we first aimed to show that the results in the case of a healthy left ventricle agree well enough with those reported in the literature, which includes \textit{in vivo} data. However, the effects of this twist in the case of aortic regurgitation may induce a spiraling component to the regurgitant and mitral jets, the resulting interaction of which remains to be explored.
		
	\section*{Acknowledgments}
		
		This work was supported by a grant from the Natural Sciences and Engineering Research Council of Canada (Grant No.\ 343164-07). G.D.L.\ was supported by the Vanier Canada Graduate Scholarship. The authors would also like to thank Dr.\ Eyal Ben-Assa for his insightful comments and suggestions as well as the undergraduate students of Team 16 in the Capstone design project at Concordia University (Montr\'{e}al, QC) for their hard work in the first and most critical stage of development of the heart duplicator used in this study, particularly Alexandre B\'{e}langer, Emilia Benevento, and Nick Ghaffari. Additionally, we owe gratitude to 3R Cardio Inc.\ for the use of their signal conditioner and fiber-optic pressure sensors in acquiring the pressure waveforms reported in this work. We also would like to thank St\'{e}fan van der Walt and Nathaniel Smith for their perceptually uniform colormap, and Ander Biguri for its MATLAB implementation, used in plotting the time-frequency spectra reported in this study as well as Adam Auton for the red-blue colormap used in plotting the vorticity contours.
		
	\section*{Supplemental Material}
	
		The reader is referred to \href{https://doi.org/10.1103/PhysRevFluids.3.113101}{\url{https://doi.org/10.1103/PhysRevFluids.3.113101}} for the supplemental material.
		
	\bibliographystyle{apacite}

\end{document}